\documentclass[final]{IEEEtran}
\pdfoutput=1
\usepackage{amsthm}
\usepackage{amsmath,epsfig,amssymb,verbatim,amsopn,subfigure,color,amsmath}
\usepackage{cite,xspace}
\usepackage{array,algorithm,algorithmic}
\usepackage{array,xspace}
\usepackage{multirow,array}%
\usepackage{multicol}%
\usepackage{dsfont,float}
\usepackage{stfloats,graphicx}
\usepackage{tikz,relsize,makecell,enumerate}

\normalsize

\ifCLASSOPTIONpeerreview
\renewcommand{\baselinestretch}{1.64}
\fi

\ifCLASSOPTIONonecolumn

 \usepackage[margin=1in]{geometry}
\else
    
\fi


\newcommand{\redcom}[1]{{\color{black}#1}\xspace}

\usepackage{relsize}
\newcommand{\densityD}{\lambda}
\newcommand{\densityR}{\lambda^{\textup{DRx}}}
\newcommand{\phiD}{\Phi^{\textup{DRx}}}
\newcommand{\phiDUE}{\Phi^{\textup{DUE}}}
\newcommand{\frd}{f_{R_d}(r_d)}
\newcommand{\frc}{f_{R_c}(r_c)}
\newcommand{\frg}{f_{G}(g)}
\newcommand{\Inftagg}{I_{\textup{agg}}}
\newcommand{\InftaggD}{I_{\textup{agg}}^{\textup{DRx}}}
\newcommand{\InftaggBS}{I_{\textup{agg}}^{\textup{BS}}}
\newcommand{\InfD}{I^{\textup{DRx}}}
\newcommand{\InfB}{I^{\textup{BS}}}
\newcommand{\BS}{\textup{BS}}
\newcommand{\DRx}{\textup{DRx}}

\newcommand{\area}{\mathcal{A}}
\newcommand{\out}{\textup{out}}
\newcommand{\DD}{\textup{D2D}}
\newcommand{\rd}{R_D}
\newcommand{\rdd}{R_{D1}}
\newcommand{\rddd}{R_{D2}}
\newcommand{\im}{\textup{\textbf{i}}}
\newcommand{\RM}{R}
\newcommand{\totalphi}{\Phi}
\newcommand{\Inftcue}{I^{\textup{DRx}}_C}
\newcommand{\dis}{d}

\ifCLASSOPTIONonecolumn
\fi

\newtheorem{corollary}{Corollary}
\newtheorem{lemma}{Lemma}
\newtheorem{prop}{Proposition}
\graphicspath{{figs/},{Figures/}}

\newcommand{\AuthorOne}{Jing~Guo, {\em{Student Member, IEEE}}}
\newcommand{\AuthorTwo}{Salman~Durrani, {\em{Senior Member, IEEE}}}
\newcommand{\AuthorFour}{Halim~Yanikomeroglu, {\em{Senior Member, IEEE}}}
\newcommand{\AuthorThree}{Xiangyun~Zhou, {\em{Member, IEEE}}}
\newcommand{\ThankOne}{J. Guo, S. Durrani and X. Zhou are with the Research School of Engineering, The Australian National University, Canberra, ACT 2601, Australia (Emails: \{jing.guo, salman.durrani, xiangyun.zhou\}@anu.edu.au). H. Yanikomeroglu is with the Department of Systems and Computer Engineering, Carleton University, Ottawa, ON K1S 5B6, Canada (E-mail: halim@sce.carleton.ca).}

\begin{document}

\title{Device-to-Device Communication Underlaying a Finite Cellular Network Region}

\author{\IEEEauthorblockN{\AuthorOne,~\AuthorTwo,~\AuthorThree~and~\AuthorFour\thanks{\ThankOne}}}

\maketitle

%
\begin{abstract}
Underlay in-band device-to-device (D2D) communication can improve the spectrum efficiency of cellular networks. However, the coexistence of D2D and cellular users causes inter-cell and intra-cell interference. The former can be effectively managed through inter-cell interference coordination \redcom{and, therefore, is not considered in this work. Instead, we focus on the intra-cell interference and} propose a D2D mode selection scheme to manage it inside a finite cellular network region. The potential D2D users are controlled by the base station (BS) to operate in D2D mode based on the average interference generated to the BS. Using stochastic geometry, we study the outage probability experienced at the BS and a D2D receiver, and spectrum reuse ratio, which quantifies the average fraction of successfully transmitting D2D users. The analysis shows that the outage probability at the D2D receiver varies for different locations. Additionally, without impairing the performance at the BS, if the path-loss exponent on the cellular link is \redcom{slightly} lower than that on the D2D link, the spectrum reuse ratio \redcom{can have negligible decrease} while the D2D users' average number of successful transmissions increases with increasing D2D node density. This indicates that an increasing level of D2D communication can be beneficial in future networks.
\end{abstract}
\begin{IEEEkeywords}
Device-to-device communication, intra-cell interference, spectrum reuse ratio, stochastic geometry, location-dependent performance.
\end{IEEEkeywords}

\ifCLASSOPTIONpeerreview
    \newpage
\fi

\section{Introduction}
\subsection{Background} Device-to-device (D2D) communication, allowing direct communication between nearby users, is envisioned as an innovative feature of 5G cellular networks~\cite{Asadi-2014,Tehrani-2014,Boccardi-2014}. Different from ad-hoc networks, the D2D communication is generally established under the control of the base station (BS). In D2D-enabled cellular networks, the cellular and D2D users can share the spectrum resources in two ways: \textit{in-band} where D2D communication utilizes the cellular spectrum and \textit{out-of-band} where D2D communication utilizes the unlicensed spectrum~\cite{outband-2015}. In-band D2D can be further divided into two categories: \textit{overlay} where the cellular and D2D communications use orthogonal (i.e., dedicated) spectrum resources and \textit{underlay} where D2D users share the same spectrum resources occupied by the cellular users. Note that the spectrum sharing in in-band D2D is controlled by the cellular network, which is different than the spectrum sharing in cognitive radio networks~\cite{Lin-m-2014,Guo-2015}. Underlay in-band D2D communication can greatly improve the spectrum efficiency of cellular networks and is considered in this paper.
\ifCLASSOPTIONpeerreview
\else
\begin{table*}[t]
\centering
\caption{Poisson Point Process and General Channel Model Variables}\label{tb:1}
\begin{tabular}{|c|c||c|c|} \hline
Symbol & Meaning & Symbol & Meaning \\ \hline
$\totalphi$ & PPP of potential D2D users (p-DUEs) & $\alpha_{\!C}$ & path-loss exponent on cellular link \\ \hline
$\phiD$ & PPP of D2D receivers (DRxs) & $\alpha_{\!D}$ & path-loss exponent on D2D link \\ \hline
\multirow{2}{*}{$\phiDUE$}& \multirow{2}{*}{\makecell[c]{PPP of DUEs\\(i.e., p-DUE in underlay D2D mode)}}& \multirow{2}{*}{$g_z$} & \multirow{2}{*}{\makecell[c]{fading power gain on the interfering link between cellular \\ users (CUE) and typical DRx link; i.i.d. Rayleigh fading}}\\
&& &\\\hline
\multirow{2}{*}{$\phiD_u$}& \multirow{2}{*}{\makecell[c]{PPP of underlay DRxs (i.e., the\\ corresponding p-DUEs in underlay D2D mode)}}& \multirow{2}{*}{$g_k^{\kappa}$} & \multirow{2}{*}{\makecell[c]{fading power gain on the interfering link between $k$-th DUE\\and typical Rx $\kappa$; i.i.d. Rayleigh fading}}\\
&& &\\\hline
$\densityD$ & Density of the PPP $\totalphi$ &$g_0$ &fading power gain on the desired link; Nakagami-$m$ fading\\ \hline
$\densityR$& Density of the PPP $\phiD$ & $\rho_{\!D}$& receiver sensitivity of DRx\\ \hline
$z$ & CUE itself and its location & $\rho_{\BS}$ & receiver sensitivity of BS\\ \hline
$x_k$ & $k$-th p-DUE itself and its location & $r_{z}$ & distance between CUE and BS\\ \hline
$y_k$ & $k$-th DRx itself and its location & $r_{c_k}$ & distance between $k$-th p-DUE and BS\\ \hline
$y'$ & typical DRx: distance $d$ away from the BS & $r_{d_k}$ & distance between $k$-th p-DUE and its DRx \\ \hline
\end{tabular}
\end{table*}
\fi
\subsection{Motivation and related work}
A key research challenge in underlay in-band D2D is how to deal with the interference between D2D users and cellular users. For traditional cellular networks with universal reuse frequency, the inter-cell interference coordination (ICIC) and its enhancements can be used to effectively manage the inter-cell interference. Thus, dealing with intra-cell interference in D2D-enabled cellular networks becomes a key issue. Existing works have proposed many different approaches to manage the interference, which have been summarized in~\cite{Asadi-2014}. The main techniques include: (i) Using network coding to mitigate interference~\cite{OSSEIRAN-2009}. However, this increases the implementation complexity at the users. (ii) Using interference aware/avoidence resource allocation methods~\cite{Janis-2009,Zhang-2013,Peng-2014,Yu-2014,Sheng-2016}. These can involve advanced mathematical techniques such as optimization theory, graph theory or game theory. (iii) Using mode selection which involves choosing to be in underlay D2D mode or not. In this regard, different mode selection schemes have been proposed and analyzed in infinite regions using stochastic geometry in~\cite{ElSawy-2014,Lin-2014,Marshall-2015,George-2015,Stefanatos-2015}. These schemes generally require knowledge of the channel between cellular and D2D users. (iv) Using other interference management techniques such as advanced receiver techniques, power control, etc.~\cite{Min-2011,Yu-2011,Ni-2015,Lee-2015,7236879}.

Since D2D communication is envisaged as short-range direct communication between nearby users, it is also very important to model the D2D-enabled cellular networks as finite regions as opposed to infinite regions. The consideration of finite regions allows modeling of the location-dependent performance of users (i.e., the users at cell-edge experience different interference compared with users in the center). In this regard it is a highly challenging open problem to analytically investigate the intra-cell interference in a D2D-enabled cellular network and the performance of underlay D2D communication when the users are confined in a finite region.


\subsection{Contributions}
In this paper, we model the cellular network region as a finite size disk region and assume that multiple D2D users are confined inside this finite region, where their locations are modeled as a Poisson Point Process (PPP). The D2D users share the uplink resources occupied by cellular users. \redcom{In this work, we do not consider the inter-cell interference and assume that it is effectively managed by the inter-cell interference coordination scheme}. Since D2D users are allowed to share the cellular user's spectrum (i.e., underlay in-band D2D paradigm), the overall network performance is governed by the intra-cell interference. Hence, we focus on the intra-cell interference in this paper. In order to ensure quality-of-service (QoS) at the BS and to manage the intra-cell interference at the BS, we consider a mode selection scheme, as inspired from~\cite{Kaufman-2008,Peng-2009}, which allows a potential D2D user to be in underlay D2D mode according to its average interference generated to the BS. In order to provide quality-of-service at the D2D users, we assume that a successful transmission occurs only if the signal-to-interference ratio (SIR) at the D2D receiver is greater than a threshold. The main contributions of this work are as follows:
\begin{itemize}
\item  Using the stochastic geometry, we derive approximate yet accurate analytical results for the outage probability at the BS and a typical D2D user, as summarized in Propositions 1 and 2, by assuming Nakagami-$m$ fading channels, a path-loss exponent of 2 or 4 for D2D link and the full channel inversion power control (i.e., the intended receiver (BS or D2D user) has the minimum required received power which is known as the receiver sensitivity). The outage probability at the D2D user highlights the location-dependent performance in a finite region.

\item Based on the derived outage probability at the D2D user, we propose and analyze two metrics to evaluate the overall quality of underlay D2D communication, namely the \textit{average number of successful D2D transmissions}, which is the average number of successful transmissions for underlay D2D users over the finite network region, and the \textit{spectrum reuse ratio} which quantifies the average fraction of underlay D2D users that can transmit successfully in the finite region. Using the derived analytical expressions, which are summarized in Propositions 1-5, we investigate the effects of the main D2D system parameters on these two metrics under the constraint of achieving certain QoS at the BS.

\item Our numerical results show that when the D2D receiver sensitivity is not too small compared to the receiver sensitivity of BS, the average number of successful D2D transmissions over the finite network area increases, while the spectrum reuse ratio decreases with increasing D2D user's node density. However, if the path-loss exponent on the cellular link is \redcom{slightly} lower than the path-loss exponent on the D2D link, \redcom{then the spectrum reuse ratio can have negligible degradation} with the increase of node density. This is important since an increasing level of D2D usage is expected in future networks and our numerical results help to identify scenarios where increasing D2D node density is beneficial to underlay D2D communications, without compromising on the cellular user's performance.
\end{itemize}

\subsection{Paper organization and notations}
The remainder of this paper is organized as follows. Section~\ref{sec:systemmodel} describes the network model and assumptions, including the mode selection scheme. Section~\ref{sec:outage} presents the analytical results for the outage probability at BS and a typical D2D receiver. Section~\ref{sec:link} proposes and derives two metrics to assess the overall quality of underlay D2D communication in a finite region. Section~\ref{sec:result} presents the numerical and simulation results, and uses the numerical results to obtain design guidelines. Finally, Section~\ref{sec:conclusion} concludes the paper.

 The following notation is used in the paper. $\Pr(\cdot)$ indicates the probability measure. $|\cdot|$ denotes the area of a certain network region and $\textup{abs}(\cdot)$ is the absolute value. \textbf{i} is the imaginary number $\sqrt{-1}$ and Im$\{\cdot\}$ denotes the imaginary part of a complex-valued number. $\textup{acos}(\cdot)$ is the inverse cosine function. $\Gamma\!\left[\cdot\right]$ is the Gamma function, while $_2F_1[\cdot,\cdot;\cdot;\cdot]$ and $\textup{MeijerG}\left[\{\cdot\},\cdot\right]$  represent the ordinary hypergeometric function and the Meijer G-function, respectively~\cite{gradshteyn2007}. Furthermore, given $f(x)$ is a function of $x$, $\left[f(x)\right]|_a^b \triangleq f(b)-f(a)$. Table~\ref{tb:1} summarizes the main PPP and channel model variables used in this paper.
\ifCLASSOPTIONpeerreview
\begin{table*}[t]
\centering
\caption{Poisson Point Process and General Channel Model Variables}\label{tb:1}
\begin{tabular}{|c|c||c|c|} \hline
Symbol & Meaning & Symbol & Meaning \\ \hline
$\totalphi$ & PPP of potential D2D users (p-DUEs) & $\alpha_{\!C}$ & path-loss exponent on cellular link \\ \hline
$\phiD$ & PPP of D2D receivers (DRxs) & $\alpha_{\!D}$ & path-loss exponent on D2D link \\ \hline
\multirow{2}{*}{$\phiDUE$}& \multirow{2}{*}{\makecell[c]{PPP of DUEs\\(i.e., p-DUE in underlay D2D mode)}}& \multirow{2}{*}{$g_z$} & \multirow{2}{*}{\makecell[c]{fading power gain on the interfering link between cellular \\ users (CUE) and typical DRx link; i.i.d. Rayleigh fading}}\\
&& &\\\hline
\multirow{2}{*}{$\phiD_u$}& \multirow{2}{*}{\makecell[c]{PPP of underlay DRxs (i.e., the\\corresponding p-DUEs in underlay D2D mode)}}& \multirow{2}{*}{$g_k^{\kappa}$} & \multirow{2}{*}{\makecell[c]{fading power gain on the interfering link between $k$-th DUE\\and typical Rx $\kappa$; i.i.d. Rayleigh fading}}\\
&& &\\\hline
$\densityD$ & Density of the PPP $\totalphi$ &$g_0$ &fading power gain on the desired link; Nakagami-$m$ fading\\ \hline
$\densityR$& Density of the PPP $\phiD$ & $\rho_{\!D}$& receiver sensitivity of DRx\\ \hline
$z$ & CUE itself and its location & $\rho_{\BS}$ & receiver sensitivity of BS\\ \hline
$x_k$ & $k$-th p-DUE itself and its location & $r_{z}$ & distance between CUE and BS\\ \hline
$y_k$ & $k$-th DRx itself and its location & $r_{c_k}$ & distance between $k$-th p-DUE and BS\\ \hline
$y'$ & typical DRx: distance $d$ away from the BS & $r_{d_k}$ & distance between $k$-th p-DUE and its DRx \\ \hline
\end{tabular}
\end{table*}
\else
\fi

\section{System Model}\label{sec:systemmodel}
Consider a single cellular network that employs the orthogonal frequency-division multiple scheme with a center-located base station. The region of cell $\area$  is assumed to be a finite disk with radius $\RM$ and area $|\area|=\pi \RM^2$. We assume that the inter-cell interference is effectively managed with ICIC mechanism, based on resource scheduling. Hence, the inter-cell interference is not considered in this paper. This assumption has been widely used in the literature, e.g., see~\cite{OSSEIRAN-2009,Zhang-2013,Yu-2014,Sheng-2016,Min-2011,Yu-2011,Ni-2015,Lee-2015,7236879}. We also restrict our analysis to one uplink channel because the other channels occupied by CUEs share similar interference statistics~\cite{ElSawy-2014,Lin-2014,Marshall-2015,George-2015,Lee-2015}. For analytical convenience, we assume that there is one cellular uplink user (CUE), whose location follows a uniform distribution inside the entire cellular region (i.e., from $0$ to $\RM$). Let $z$ denote both the location of the CUE and the cellular user itself.
\begin{figure}[t]
\centering
\begin{tikzpicture}[scale=0.3 ]
\tikzstyle{every node}=[font=\small]
\draw  (0,0) circle (8);
\fill[red] (0,0.5) -- (-0.5,-0.55) -- (0.65,-0.5) -- (0,0.5);
\draw[thick,red]  (2,4.05)circle( 0.5);

\fill[blue] (-7,-1) rectangle (-6.15,-1.8);
\fill[blue]  (-4,-3)circle(0.5);
\draw[dashed,blue]  (-4,-3)--(-6.25,-1.95);

\draw[blue,thick]  (-2,2.45) rectangle (-1.1,1.5);
\fill[blue,thick]  (-2,5) circle (0.5);
\draw[dashed,blue] (-1.85,4.35) -- (-1.5,2.5);

\fill[blue]  (2,-5.75) rectangle (1,-4.7);
\fill[blue] (4,-6.25)circle( 0.5);
\draw[dashed,blue] (4,-6.3) -- (1.5,-5.25);

\draw (1.15,-5.45)-- (-4,-3);
\draw (-4,-3)--(-0.05,-0.15);
\draw (1.25,-4.7)--(-0.05,-0.15);

\draw(1.5,-3) node[above]{$r_c$};
\draw(3,-6.5) node[above]{$r_d$};
\draw(-2.5,-3.5) node[above]{$\theta$};
\draw(-3,-1.5) node[above]{$\dis$};
\draw[dashed,->] (0,0) -- (8,0);
\draw(3.5,-0.5) node[above]{$\RM$};
\end{tikzpicture}
               \vspace{-0.05 in}
\caption{Illustration of the network model (${\color{red}\blacktriangle}=$ BS, ${\color{red}\circ}=$ CUE, ${\color{blue}\blacksquare}=$ DUE (p-DUE in D2D mode), ${\color{blue}\square}=$ p-DUE in other transmission mode, ${\color{blue}\bullet}=$ DRx. Note that p-DUE and its corresponding DRx are connected by a dashed line). }\label{fig:system}
\vspace{-0.05 in}
\end{figure}

To improve the spectral efficiency of the frequency band occupied by the CUE, its uplink channel is also utilized for D2D communication. Note that D2D communication may also reuse downlink resources, but uplink is preferred in terms of interference in practical systems as it is less congested~\cite{Lin-m-2014}. We further assume there are multiple potential D2D users (p-DUEs) that are randomly distributed in the entire cellular region $\area$. Note that the distributions of CUE and p-DUE are assumed to be independent. For each p-DUE, there is an intended receiver (DRx) which is uniformly distributed within this p-DUE's proximity (e.g., $\pi \rd^2$)\footnote{In reality, the intended DRx should also be confined in the network region $\area$ (i.e., a disk region with radius $\RM$). However, for p-DUE nears cell-edge, this would mean that the DRx is no longer uniformly distributed in a disk region. For analytical tractability, we still assume that DRx is uniformly distributed in a disk region, regardless of the p-DUE's location, i.e., we assume that the DRx is confined in a disk region of radius $\RM+\rd$. The accuracy of this approximation will be validated in the results section.}, hence, the distance distribution for the potential D2D link, $r_d$, is $\frd=\frac{2r_d}{\rd^2}$. Let $x_k$ denote both the location of the $k$-th p-DUE and the user itself, and $y_k$ denote both the location of the $k$-th DRx and the receiver itself. For analytical convenience, we further assume that the location of p-DUE follows the Poisson Point Process, denoted as $\totalphi$, with constant density $\densityD$. Thus, based on the displacement property of PPP~\cite[eq. (2.9)]{Haenggi-2012}, the location of DRxs also follows a PPP, denoted as $\phiD$, with density $\densityR$.

We consider the path-loss plus block fading channel model. In this way, the received power at a receiver (Rx) is $P_t g r^{-\alpha}$, where $P_t$ is the transmit power of the transmitter, $g$ denotes the fading power gain on the link that is assumed to be independently and identically distributed (i.i.d.), $r$ is the distance between the transmitter and receiver, and $\alpha$ is the path-loss exponent. Additionally, as we consider the uplink transmission, power control is necessary; we employ the full channel inversion for uplink power control~\cite{Lin-2014,ElSawy-2014}. Hence, the transmit power for the CUE and the p-DUE using D2D link are $\rho_{\BS} r_z^{\alpha_{\!C}}$ and $\rho_{\!D} r_d^{\alpha_{\!D}}$, respectively, where $r_z$ is the distance between CUE and BS, $\rho_{\BS}$ and $\rho_{\!D}$ are the minimum required power at BS and DRx (also known as the receiver sensitivity), and $\alpha_{\!C}$ and $\alpha_{\!D}$ are path-loss exponents on cellular link and D2D link, respectively.

\redcom{We define the mode selection scheme as the selection between the underlay D2D mode (i.e., direct communication via the D2D link in underlay paradigm) or the other transmission mode. The other transmission mode can be the overlay D2D mode where the dedicated spectrum that is not occupied by cellular user is used, or the silent mode where no transmission happens~\cite{6364738,Phunchongharn-2013,Mach-2015}.} In this paper, as motivated by~\cite{Kaufman-2008,Peng-2009}, we consider that the mode selection for each p-DUE is determined by its average interference generated to the BS. For example, if the average interference $\rho_{\!D} r_{d_k}^{\alpha_{\!D}}r_{c_k}^{-\alpha_{\!C}}$ for the $k$-th p-DUE is larger than the threshold $\xi$, where $r_{d_k}$ is the distance between this p-DUE and BS, then, this user is forced by the BS to operate in the other transmission mode. The focus of this paper is on potential D2D users in underlay in-band D2D mode (referred to as DUEs which follow a certain point process $\phiDUE$); the analysis of the other transmission mode is outside the scope of this work. Also we assume that the BS is fully in control of the D2D communication and D2D device discovery, which ensures that the considered mode selection scheme is feasible~\cite{Mach-2015}.

In the above set-up, intra-cell interference exists in the network because the considered spectrum band is shared between a CUE and DUEs. The aggregate interference received at the BS and at a certain DRx $y_j$ can then be expressed as
\ifCLASSOPTIONonecolumn
 \begin{subequations}\label{eq:Iagg_lterm}
\begin{align}
\InftaggBS
&=\sum\limits_{x_{k}\in\totalphi}g_k^{\BS}\rho_{\!D} r_{d_k}^{\alpha_{\!D}}r_{c_k}^{-\alpha_{\!C}}\textbf{1}\!\left(\rho_{\!D} r_{d_k}^{\alpha_{\!D}}r_{c_k}^{-\alpha_{\!C}}\!<\!\xi\right),\label{eq:IaggCUE_lterm}\\
\InftaggD(x_j,y_j)
&=\frac{g_z\rho_{\BS}r_z^{\alpha_{\!C}}}{|z-y_j|^{\alpha_{\!D}}}+\sum\limits_{x_{k}\in\totalphi,k\neq j}\!\frac{g_k^{\DRx}\rho_{\!D}r_{d_k}^{\alpha_{\!D}}}{|x_k-y_j|^{\alpha_{\!D}}}\textbf{1}\!\left(\rho_{\!D} r_{d_k}^{\alpha_{\!D}}r_{c_k}^{-\alpha_{\!C}}\!<\!\xi\right),\label{eq:IaggDRx_lterm}
\end{align}
\end{subequations}
\else
 \begin{subequations}\label{eq:Iagg_lterm}
\begin{align}
\InftaggBS
&=\sum\limits_{x_{k}\in\totalphi}g_k^{\BS}\rho_{\!D} r_{d_k}^{\alpha_{\!D}}r_{c_k}^{-\alpha_{\!C}}\textbf{1}\!\left(\rho_{\!D} r_{d_k}^{\alpha_{\!D}}r_{c_k}^{-\alpha_{\!C}}\!<\!\xi\right),\label{eq:IaggCUE_lterm}\\
\InftaggD(x_j,y_j)
&=\frac{g_z\rho_{\BS}r_z^{\alpha_{\!C}}}{|z-y_j|^{\alpha_{\!D}}}\nonumber\\
&+\sum\limits_{x_{k}\in\totalphi,k\neq j}\!\frac{g_k^{\DRx}\rho_{\!D}r_{d_k}^{\alpha_{\!D}}}{|x_k-y_j|^{\alpha_{\!D}}}\textbf{1}\!\left(\rho_{\!D} r_{d_k}^{\alpha_{\!D}}r_{c_k}^{-\alpha_{\!C}}\!<\!\xi\right),\label{eq:IaggDRx_lterm}
\end{align}
\end{subequations}
\fi
respectively, where $\textbf{1}(\cdot)$ is the indicator function, $|z-y_j|$ and $|x_k-y_j|$ denote the Euclidean distance between CUE and $j$-th DRx, $k$-th DUE and $j$-th DRx, respectively. $g_z$, $g_k^{\BS}$ and $g_k^{\DRx}$ are the fading power gain on the interfering links, which are assumed to be the i.i.d. Rayleigh fading. In the following, we refer $\Inftagg^{\kappa}$ to the aggregate interference at a typical Rx $\kappa$ for notation simplicity.

Considering an interference-limited system, we can write the signal-to-interference ratio at a typical Rx $\kappa$ as
\begin{align}
\textup{SIR}^{\kappa} &=\frac{g_0\rho}{\Inftagg^{\kappa}},
\end{align}
where $g_0$ is the fading power gain on the reference link between the typical transmitter-receiver pair, which is assumed to exprience Nakagami-$m$ fading, $\rho$ is the receiver sensitivity of the typical Rx (i.e., $\rho=\rho_{\BS}$ when BS is the typical Rx and $\rho=\rho_{\!D}$ if DRx is the typical Rx)\footnote{According to~\eqref{eq:IaggDRx_lterm}, when the typical Rx is a DRx, the SIR relies on the location of DRx $y_j$. Hence, the SIR at $y_j$ should be expressed as $\textup{SIR}^{\DRx}(x_j,y_j)$. But in this paper, we will sometimes ignore $x_j$ and $y_j$, and refer it simply as $\textup{SIR}^{\DRx}$. This notation is also adopted for the outage probability at $y_j$, where the full notation would be $P_{\out}^{\kappa}(\gamma,y_j)$. }.
\section{Outage Probability Analysis}\label{sec:outage}
To evaluate the network performance, we first consider and compute the outage probability experienced at a typical receiver.

\subsection{Mathematical framework}
Our considered outage probability for a typical Rx at a given location is averaged over the fading power gain and the possible locations of all interfering users. Mathematically, the outage probability at a typical Rx can be written as
\begin{align}\label{eq:outage}
P_{\out}^{\kappa}(\gamma)=\mathbb{E}_{\Inftagg^{\kappa},g_0}\left\{\Pr\left(\frac{g_0\rho}{\Inftagg^{\kappa}}<\gamma\right)\right\},
\end{align}
\noindent where $\mathbb{E}_{\Inftagg^{\kappa},g_0}\left\{\cdot\right\}$ is the expectation operator with respect to $\Inftagg^{\kappa}$ and $g_0$.

We leverage the reference link power gain-based framework~\cite{Guo-2013} to work out the outage probability. For the case that the reference link suffers from the Nakagami-m fading with integer $m$, the outage expression in~\eqref{eq:outage} can be rewritten as (see proof in Appendix~\ref{app_outage})
\begin{align}\label{eq:outage_naka}
P_{\out}^{\kappa}(\gamma)=&1-\sum\limits_{t=0}\limits^{m-1}\frac{(-s)^{t}}{t!}\left.\frac{d^t}{d s^t}\mathcal{M}_{\Inftagg^{\kappa}}(s)\right|_{s=m\frac{\gamma}{\rho}},
\end{align}
\noindent where $\mathcal{M}_{\Inftagg^{\kappa}}(s)=\mathbb{E}_{\Inftagg^{\kappa}}\left[\exp(-s\Inftagg^{\kappa})\right]$ is the moment generating function (MGF) of $\Inftagg^{\kappa}$. Note that this fading model covers Rayleigh fading (i.e., by setting $m=1$) and can also approximate Rician fading~\cite{Guo-2013,Marvin2005}. Hence, it is adopted in this paper.

As shown in~\eqref{eq:outage_naka}, the computation of outage probability requires the MGF results for the aggregate interference at the typical Rx, which will be presented in the following.

\subsection{MGF of the aggregate interference at the BS}
The aggregate interference at the BS is generally in the form of $\sum\limits_{x_{k}\in\totalphi}\InfB_{k}$, where $\InfB_{k}$ is the interference from the $k$-th p-DUE. Note that $\InfB_{k}=0$ if $k$-th p-DUE is in the other transmission mode. Due to the independently and uniformly distributed (i.u.d.) property of p-DUEs and the i.i.d. property of the fading channels, the interference from a p-DUE is also i.i.d.. In the following, we drop the index $k$ in $r_{c_k}$, $r_{d_k}$, $g_k$ and $\InfB_{k}$. As such, the aggregate interference can be written as $\left(\InfB\right)^M$, where $M$ is the number of p-DUEs following the Poisson distribution with density $\lambda(|\area|)$. Based on the MGF's definition (stated below \eqref{eq:outage_naka}), the MGF of $\InftaggBS$ is given by
\ifCLASSOPTIONonecolumn
\begin{align}\label{eq:MGFTotalbs}
\mathcal{M}_{\InftaggBS}(s)=\mathbb{E}_M\left[\left.\mathbb{E}_{\InfB}\left[\exp\left(-s\left(\InfB\right)^M\right)\right|M\right]\right]=\exp\Big(\densityD(|\area|)\big(\mathcal{M}_{\InfB}(s)-1\big)\Big),
\end{align}
\else
\begin{align}\label{eq:MGFTotalbs}
\mathcal{M}_{\InftaggBS}(s)=&\mathbb{E}_M\left[\left.\mathbb{E}_{\InfB}\left[\exp\left(-s\left(\InfB\right)^M\right)\right|M\right]\right]\nonumber\\
&=\exp\Big(\densityD(|\area|)\big(\mathcal{M}_{\InfB}(s)-1\big)\Big),
\end{align}
\fi
\noindent where $\mathcal{M}_{\InfB}(s)$ denotes the MGF of the interference at the BS from a p-DUE, which is presented as follows.

\begin{prop}\label{prop_schemeBS}
\ifCLASSOPTIONonecolumn
For the underlay in-band D2D communication with the considered mode selection scheme in a disk-shaped cellular network region, following the system model in Section~\ref{sec:systemmodel}, the MGF of the interference from an i.u.d. p-DUE received at the BS can be expressed as
  \begin{align}\label{eq:MGF_BS_lt_final}
\mathcal{M}_{\InfB}(s)
 =&1+\!\frac{\,_2F_1\left[1,\frac{2}{\alpha_{\!C}};1+\frac{2}{\alpha_{\!C}};\frac{-1}{s\xi}\right]}{\rd^2\RM^2\tilde{\rd}^{\!\!-2-\frac{2\alpha_{\!D}}{\alpha_{\!C}}}(\xi/\rho_{\!D})^{\frac{2}{\alpha_{\!C}}}}\frac{\alpha_{\!C}}{\alpha_{\!D}+\alpha_{\!C}}\nonumber\\
 &\quad-\begin{cases}
\left.\left[\frac{\!\alpha_{\!C} \,_2F_1\left[1,\frac{2}{\alpha_{\!C}};1+\frac{2}{\alpha_{\!C}};\frac{-\RM^{\alpha_{\!C}}}{s\rho_{\!D} x^{\alpha_{\!D}}}\right]
\!+\!\alpha_{\!D} \,_2F_1\!\left[1,\frac{-2}{\alpha_{\!D}};1-\frac{2}{\alpha_{\!D}};\frac{-\RM^{\alpha_{\!C}}}{s\rho_{\!D} x^{\alpha_{\!D}}}\right]}{x^{-2}\rd^2(\alpha_{\!C}+\alpha_{\!D})}\right]\right|_0^{\tilde{\rd}}
,  &{\alpha_{\!D}\neq2}; \\
\frac{2\tilde{\rd}^2\textup{MeijerG}\left[\left\{\left\{0,\frac{\alpha_{\!C}-2}{\alpha_{\!C}}\right\},\{2\}\right\}, \left\{\left\{0,1\right\},\left\{\frac{-2}{\alpha_{\!C}}\right\}\right\},\frac{\RM^{\alpha_{\!C}}}{s\rho_{\!D}\tilde{\rd}^2}\right]}{\rd^2\alpha_{\!C}}
,  &{\alpha_{\!D}=2};
\end{cases}
  \end{align}
  \noindent where $\tilde{\rd}\triangleq\textup{min}\left(\rd,\RM^{\frac{\alpha_{\!C}}{\alpha_{\!D}}}\left(\frac{\xi}{\rho_{\!D}}\right)^{\frac{1}{\alpha_{\!D}}}\right)$ and $\xi$ is the mode selection threshold.
  \else
    \begin{figure*}[!t]
\normalsize
\begin{align}\label{eq:MGF_BS_lt_final}
\mathcal{M}_{\InfB}(s)
 =&1+\!\frac{\,_2F_1\left[1,\frac{2}{\alpha_{\!C}};1+\frac{2}{\alpha_{\!C}};\frac{-1}{s\xi}\right]}{\rd^2\RM^2\tilde{\rd}^{\!\!-2-\frac{2\alpha_{\!D}}{\alpha_{\!C}}}(\xi/\rho_{\!D})^{\frac{2}{\alpha_{\!C}}}}\frac{\alpha_{\!C}}{\alpha_{\!D}+\alpha_{\!C}}-\!\begin{cases}
\!\!\left.\left[\frac{\!\alpha_{\!C} \,_2F_1\left[1,\frac{2}{\alpha_{\!C}};1+\frac{2}{\alpha_{\!C}};\frac{-\RM^{\alpha_{\!C}}}{s\rho_{\!D} x^{\alpha_{\!D}}}\!\right]
\!+\!\alpha_{\!D} \,_2F_1\!\left[1,\frac{-2}{\alpha_{\!D}};1-\frac{2}{\alpha_{\!D}};\frac{-\RM^{\alpha_{\!C}}}{s\rho_{\!D} x^{\alpha_{\!D}}}\!\right]}{x^{-2}\rd^2(\alpha_{\!C}+\alpha_{\!D})}\!\right]\!\right|_0^{\tilde{\rd}}
,  &{\alpha_{\!D}\neq2}; \\
\frac{2\tilde{\rd}^2\textup{MeijerG}\left[\left\{\left\{0,\frac{\alpha_{\!C}-2}{\alpha_{\!C}}\right\},\{2\}\right\}, \left\{\left\{0,1\right\},\left\{\frac{-2}{\alpha_{\!C}}\right\}\right\},\frac{\RM^{\alpha_{\!C}}}{s\rho_{\!D}\tilde{\rd}^2}\right]}{\rd^2\alpha_{\!C}}
,  &{\alpha_{\!D}=2};
\end{cases}
  \end{align}
\hrulefill
\vspace*{4pt}
\vspace{-0.15 in}
\end{figure*}
  For the underlay in-band D2D communication with the considered mode selection scheme in a disk-shaped cellular network region, following the system model in Section~\ref{sec:systemmodel}, the MGF of the interference from an i.u.d. p-DUE received at the BS can be expressed as~\eqref{eq:MGF_BS_lt_final}, as shown at the top of next page, where $\tilde{\rd}\triangleq\textup{min}\left(\rd,\RM^{\frac{\alpha_{\!C}}{\alpha_{\!D}}}\left(\frac{\xi}{\rho_{\!D}}\right)^{\frac{1}{\alpha_{\!D}}}\right)$ and $\xi$ is the mode selection threshold.
\fi

\noindent Proof: See Appendix~\ref{app_scheme_BS}.
\end{prop}

Note that the result in \eqref{eq:MGF_BS_lt_final} is expressed in terms of the ordinary hypergeometric function and the MeijerG function, which are readily available in standard mathematical packages such as Mathematica.

 \subsection{MGF of the aggregate interference at a typical DRx}\label{sec:MGF-DRx}
 The point process of DUEs $\phiDUE$ is in fact an independent thinning process of the underlaying PPP $\totalphi$, which is also a PPP with a certain density~\cite{Haenggi-2012}. Similarly, in terms of the location of underlay DRxs (i.e., whose corresponding p-DUE is in underlay D2D mode), it is also a PPP $\phiD_u$, which is an independent thinning process of DRxs $\phiD$.

 For analytical convenience, we condition on an underlay DRx $y'$, which is located at a distance $\dis$ away from the BS, and its corresponding DUE is denoted as $x'$. Because of the isotropic network region and PPP's rotation-invariant property, the outage probability derived at $y'$ is the same for those underlay DRxs whose distance to BS is $\dis$. Then, according to the Slivnyak's Theorem, we can have the MGF of the aggregate interference received at $y'$ as
\ifCLASSOPTIONonecolumn
  \begin{align}\label{eq:MGFTotalDRx}
\mathcal{M}_{\InftaggD}(s,\dis)&=\mathbb{E}_{\Inftcue}\left\{\exp\left(-s\Inftcue\right)\right\}\mathbb{E}_{\totalphi\backslash x'}\left\{\exp\left(\!\!-s\!\!\sum\limits_{x_{k}\in\totalphi \backslash x'}\!\!\InfD_k(y')\right)\right\}\nonumber\\
&=\mathcal{M}_{\Inftcue}(s,\dis)\mathbb{E}_{\totalphi}\left\{\exp\left(\!\!-s\!\!\sum\limits_{x_{k}\in\totalphi}\!\!\InfD_k(y')\right)\right\}\nonumber\\
&=\mathcal{M}_{\Inftcue}(s,\dis)\exp\Big(\densityD(|\area|)\big(\mathcal{M}_{\InfD}(s,\dis)-1\big)\Big),
\end{align}
\else
  \begin{align}\label{eq:MGFTotalDRx}
\mathcal{M}_{\InftaggD}&(s,\dis)=\mathbb{E}_{\Inftcue}\left\{\exp\left(-s\Inftcue\right)\right\}\nonumber\\
&\quad\quad\quad\times\mathbb{E}_{\totalphi\backslash x'}\left\{\exp\left(\!\!-s\!\!\sum\limits_{x_{k}\in\totalphi \backslash x'}\!\!\InfD_k(y')\right)\right\}\nonumber\\
=&\mathcal{M}_{\Inftcue}(s,\dis)\mathbb{E}_{\totalphi}\left\{\exp\left(\!\!-s\!\!\sum\limits_{x_{k}\in\totalphi}\!\!\InfD_k(y')\right)\right\}\nonumber\\
=&\mathcal{M}_{\Inftcue}(s,\dis)\exp\Big(\densityD(|\area|)\big(\mathcal{M}_{\InfD}(s,\dis)-1\big)\Big),
\end{align}
\fi
\noindent where $\Inftcue$ is the interference from CUE, $\mathcal{M}_{\Inftcue}(s,\dis)$ is the corresponding MGF, $\InfD_k(y')$ is the interference from $k$-th p-DUE, and $\mathcal{M}_{\InfD}(s,\dis)$ is the MGF of the interference from a p-DUE. The results for these two MGFs are presented as follows.

\begin{prop}\label{prop_scheme2}
 For the underlay in-band D2D communication with the considered mode selection scheme in a disk-shaped cellular network region, following the system model in Section~\ref{sec:systemmodel}, with the path-loss exponent $\alpha_{\!C}=\alpha_{\!D}=2$ or $4$, the MGF of the interference from an i.u.d. p-DUE received at a DRx, which is a distance $\dis$ away from the BS can be given as~\eqref{eq:MGF_DRx_lt_final}, as shown at the top of next page,
   \begin{figure*}[!t]
\normalsize
\begin{align}\label{eq:MGF_DRx_lt_final}
\mathcal{M}_{\InfD}(s,\dis)=1\!-\!\begin{cases}
\frac{s\rho_{\!D}\left.\left[\Psi_1\left(x^2,s\rho_{\!D},\RM^2-\dis^2,4\dis^2s\rho_{\!D}\right)-\Psi_1\left(x^2,s\rho_{\!D}+\frac{\rho_{\!D}}{\xi},-\dis^2,4\dis^2s\rho_{\!D}\right)\right]\right|_{0}^{\tilde{\rd}}}{\rd^2\RM^2}\!, &{\alpha_{\!C}=\alpha_{\!D}=2};\\
\frac{\textup{Im}\left\{\!\left.\left[\!\Psi_1\!\left(x^2,-\im\sqrt{s\rho_{\!D}},\RM^2\!-\!\dis^2,-4\im\sqrt{s\rho_{\!D}}\dis^2\right)
\!-\!\Psi_1\!\left(x^2,\sqrt{\frac{\rho_{\!D}}{\xi}}-\im\sqrt{s\rho_{\!D}},-\!\dis^2,-4\im\sqrt{s\rho_{\!D}}\dis^2\right)\!\right]\right|_{0}^{\tilde{\rd}}\!\right\}}{(\sqrt{s\rho_{\!D}})^{-1}\rd^2\RM^2},&{\alpha_{\!C}=\alpha_{\!D}=4;}
\end{cases}
 \end{align}
\hrulefill
\vspace*{4pt}
\vspace{-0.1 in}
\end{figure*}
\noindent where $\Psi_1(x,\cdot,\cdot,\cdot)$ is given in~\eqref{eq:psi1}. Note that for other $\alpha_{\!C}$ values, the semi-closed-form of $\mathcal{M}_{\InfD}(s,\dis)$ is available in~\eqref{eq:mgf_lterm_other2} ($\alpha_{\!D}=2$) and~\eqref{eq:mgf_lterm_other4} ($\alpha_{\!D}=4$).

\noindent Proof: See Appendix~\ref{app_scheme2}.
\end{prop}

\begin{corollary}\label{prop_CUE}
\ifCLASSOPTIONonecolumn
For the underlay in-band D2D communication with the considered mode selection scheme in a disk-shaped cellular network region, following the system model in Section~\ref{sec:systemmodel}, with the path-loss exponent $\alpha_{\!C}=\alpha_{\!D}=2$ or $4$, the MGF of the interference from an i.u.d. cellular user received at a DRx, which is distance $\dis$ away from the BS, can be given as
 \begin{align}\label{eq:MGF_DRx_CUE_final}
\mathcal{M}_{\Inftcue}(s,\dis) =1-\begin{cases}
             \frac{s\rho_{\BS}\left.\left[\beta_2\left(x^2,(s\rho_{\BS}+1)^2,\dis^2(s\rho_{\BS}-1),4\dis^4s\rho_{\BS}\right)\right]\right|_{0}^{\RM}}{\RM^2(s\rho_{\BS}+1)^3}, &{\alpha_{\!C}=\alpha_{\!D}=2}; \\
             \textup{Im}\left\{\frac{\sqrt{s\rho_{\BS}}\left[\left.\beta_2\left(x^2,1-\im\sqrt{s\rho_{\BS}},-\dis^2\frac{1+\im\sqrt{s\rho_{\BS}}}{1-\im\sqrt{s\rho_{\BS}}},\frac{-4\im\sqrt{s\rho_{\BS}}\dis^4}{(1-\im\sqrt{s\rho_{\BS}})^2}\right)\right]\right|_{0}^{\RM}}{\RM^2(1-\im\sqrt{s\rho_{\BS}})^2}\right\}, &{\alpha_{\!C}=\alpha_{\!D}=4};
             \end{cases}
 \end{align}
\else
For the underlay in-band D2D communication with the considered mode selection scheme in a disk-shaped cellular network region, following the system model in Section~\ref{sec:systemmodel}, with the path-loss exponent $\alpha_{\!C}=\alpha_{\!D}=2$ or $4$, the MGF of the interference from an i.u.d. cellular user received at a DRx, which is distance $\dis$ away from the BS, can be given as~\eqref{eq:MGF_DRx_CUE_final}, as shown at the top of next page,
  \begin{figure*}[!t]
\normalsize
\vspace{-0.05 in}
 \begin{align}\label{eq:MGF_DRx_CUE_final}
\mathcal{M}_{\Inftcue}(s,\dis) =1-\begin{cases}
             \frac{s\rho_{\BS}\left.\left[\beta_2\left(x^2,(s\rho_{\BS}+1)^2,\dis^2(s\rho_{\BS}-1),4\dis^4s\rho_{\BS}\right)\right]\right|_{0}^{\RM}}{\RM^2(s\rho_{\BS}+1)^3}, &{\alpha_{\!C}=\alpha_{\!D}=2}; \\
             \textup{Im}\left\{\frac{\sqrt{s\rho_{\BS}}\left[\left.\beta_2\left(x^2,1-\im\sqrt{s\rho_{\BS}},-\dis^2\frac{1+\im\sqrt{s\rho_{\BS}}}{1-\im\sqrt{s\rho_{\BS}}},\frac{-4\im\sqrt{s\rho_{\BS}}\dis^4}{(1-\im\sqrt{s\rho_{\BS}})^2}\right)\right]\right|_{0}^{\RM}}{\RM^2(1-\im\sqrt{s\rho_{\BS}})^2}\right\}, &{\alpha_{\!C}=\alpha_{\!D}=4};
             \end{cases}
 \end{align}
\hrulefill
\vspace*{4pt}
\vspace{-0.1 in}
\end{figure*}
\fi
 \noindent where $\beta_2(x,a,b,c)=\sqrt{(ax+b)^2+c}-b\ln\left(ax+b+\sqrt{(ax+b)^2+c}\right)$. For other $\alpha_{\!C}$ values, the $\mathcal{M}_{\Inftcue}(s,\dis)$ expression is given in \eqref{eq:mgf_cue_2}~($\alpha_{\!D}=2$) and~\eqref{eq:mgf_cue_4} ($\alpha_{\!D}=4$).

\noindent Proof: See Appendix~\ref{app_coro1}.

\end{corollary}

Note that although \eqref{eq:MGF_DRx_lt_final} and \eqref{eq:MGF_DRx_CUE_final} contain terms with the imaginary number, the MGF results are still real because of the $\textup{Im}(\cdot)$ function.

\section{D2D communication performance analysis}\label{sec:link}
Generally, the outage probability reflects the performance at a typical user.  In order to characterize the overall network performance, especially when the users are confined in a finite region, metrics other than the outage probability need to be considered. In this section, we consider two metrics: average number of successful D2D transmissions and spectrum reuse ratio. Their definitions and formulations are presented below.

\subsection{Average number of successful D2D transmissions}

\subsubsection{Mathematical framework}\label{subsec:spectraleff}
In this paper, the average number of successful D2D transmissions is defined as the average number of underlay D2D users that can transmit successfully over the network region $\area$. Therein, the successful transmission is defined as the event that the SIR at a DRx is greater than the threshold $\gamma$. For the considered scenario, we obtain the expression of the average number of success transmissions in the following.

\begin{prop}\label{prop:link}
For the underlay in-band D2D communication with the considered mode selection scheme in a disk-shaped cellular network region, following the system model in Section~\ref{sec:systemmodel}, the average number of successful D2D transmissions is
 \begin{align}\label{eq:spectral2}
\bar{M}=\int_0^{\RM+\rd}\!\!\!\left(1-P_{\out}^{\DRx}(\gamma,\dis)\right) p_{\DD}(\dis)\densityR(\dis)2\pi \dis \,\textup{d}\dis,
\end{align}
\noindent where $p_{\DD}(\dis)$ is the probability that p-DUE is in D2D mode given its corresponding DRx's distance to BS is $\dis$, $\densityR(\dis)$ is the node density of DRxs, and $P_{\out}^{\DRx}(\gamma,\dis)$ is outage probability at the corresponding DRx.

\noindent Proof: See Appendix~\ref{app_link}.
\end{prop}

According to Proposition~\ref{prop:link}, the average number of successful D2D transmissions is determined by the outage probability experienced at the underlay DRxs, the density function of DRx, and the probability that the DRx is an underlay DRx. The outage probability has been derived in Section~\ref{sec:outage}. In this section, we present the results for the remaining two factors, which will then allow the computation of average number of successful D2D transmissions using~\eqref{eq:spectral2}.

\subsubsection{Density function of DRxs}

 Before showing the exact density function, we define one lemma as follows.
\begin{lemma}
For two disk regions with radii $r_1$ and $r_2$, respectively, which are separated by distance $d$, the area of their overlap region is given by~\cite{Eric-wolframe,Huang-2014}
\ifCLASSOPTIONonecolumn
\begin{align}\label{eq:psi}
\psi(d,r_1,r_2)=r_1^2\textup{acos}\!\left(\frac{d^2\!+\!r_1^2\!-\!r_2^2}{2dr_1}\right)+r_2^2\textup{acos}\!\left(\frac{d^2\!+\!r_2^2\!-\!r_1^2}{2dr_2}\right)-\frac{\sqrt{2r_2^2(r_1^2\!+\!d^2)\!-\!r_2^2\!-\!(r_1^2\!-\!d^2)^2}}{2}.
\end{align}
\else
\begin{align}\label{eq:psi}
\psi(d,r_1,r_2)=&r_1^2\textup{acos}\!\left(\frac{d^2\!+\!r_1^2\!-\!r_2^2}{2dr_1}\right)+r_2^2\textup{acos}\!\left(\frac{d^2\!+\!r_2^2\!-\!r_1^2}{2dr_2}\right)\nonumber\\
&-\frac{\sqrt{2r_2^2(r_1^2\!+\!d^2)\!-\!r_2^2\!-\!(r_1^2\!-\!d^2)^2}}{2}.
\end{align}
\fi
\end{lemma}
Using Lemma 1, we can express the node density of DRxs as shown in the following proposition.
\begin{prop}\label{theorem_density}
For a disk-shaped network region with radius $\RM$, assume that there are multiple p-DUEs that are randomly independently and uniformly distributed inside the region, and their location is modeled as a PPP with density $\densityD$. For each p-DUE, there is an intended DRx which is uniformly distributed inside the disk region formed around the p-DUE with radius $\rd$. Then, the location of DRxs also follows a PPP, with the density
  \begin{align}\label{eq:densityfunction}
 \densityR(\dis)=   \begin{cases}
             \densityD, & {0\leq \dis<\RM-\rd}; \\
             \densityD\frac{\psi(\dis,\RM,\rd)}{\pi\rd^2}, & {\RM-\rd\leq \dis\leq\RM+\rd};
             \end{cases}  %
  \end{align}
\noindent where $\psi(\cdot,\cdot,\cdot)$ is defined in Lemma 1.

Proof: See Appendix~\ref{app_Density}.
\end{prop}

The node density result in~\eqref{eq:densityfunction} can in fact be applied to a broader class of networks adopting the Poisson bi-polar network model. To the best of our knowledge, this result for the node density of receivers for the bi-polar network model in a disk region has not been presented before in the literature.


\subsubsection{Probability of being in D2D mode}\label{theorem_active2}

\begin{prop}\label{prop_active}
For the underlay in-band D2D communication with the considered mode selection scheme in a disk-shaped cellular network region, following the system model in Section~\ref{sec:systemmodel}, when the path-loss exponents for cellular link and D2D link are the same, the probability that a p-DUE is in underlay D2D mode given that its DRx's distance to BS is $\dis$, is given by
\ifCLASSOPTIONonecolumn
      \begin{align}\label{eq:pactive_2}
p_{\DD}(\dis)=   \begin{cases}
            \textbf{1}\!(\xi\!>\!\rho_{\!D})- \frac{\xi^{\frac{2}{\alpha}}\dis^2(-1)^{\textbf{1}(\xi>\rho_{\!D})+1}}{\left(\xi^{\!\frac{1}{\alpha}}-\rho_{\!D}^{\!\frac{1}{\alpha}}\right)^{\!2}\rd^2}, & {0\leq \dis<\rdd}; \\
             \textbf{1}\!(\xi\!>\!\rho_{\!D})-\frac{\psi\left(\textup{abs}\!\left(\!\frac{\xi^{\frac{2}{\alpha}}\dis}{\xi^{\frac{2}{\alpha}}-\rho_{\!D}^{\frac{2}{\alpha}}}\!\right),
             \rd,\textup{abs}\!\left(\!\frac{\xi^{\frac{1}{\alpha}}\rho_{\!D}^{\frac{1}{\alpha}}\dis}{\xi^{\frac{2}{\alpha}}-\rho_{\!D}^{\frac{2}{\alpha}}}\!\right)\right)}{(-1)^{\textbf{1}\!(\xi>\rho_{\!D})+1}\pi\rd^2}, & {\rdd\leq \dis<\rddd}; \\
             1, &{\dis\geq \rddd};
             \end{cases}  %
  \end{align}
  \else
      \begin{align}\label{eq:pactive_2}
p_{\DD}(\dis)=   \begin{cases}
            \textbf{1}\!(\xi\!>\!\rho_{\!D})- \frac{\xi^{\frac{2}{\alpha}}\dis^2(-1)^{\textbf{1}(\xi>\rho_{\!D})+1}}{\left(\xi^{\!\frac{1}{\alpha}}-\rho_{\!D}^{\!\frac{1}{\alpha}}\right)^{\!2}\rd^2}, \quad{0\leq \dis<\rdd}; \\
             \textbf{1}\!(\xi\!>\!\rho_{\!D})-\frac{\psi\left(\textup{abs}\!\left(\!\frac{\xi^{\frac{2}{\alpha}}\dis}{\xi^{\frac{2}{\alpha}}-\rho_{\!D}^{\frac{2}{\alpha}}}\!\right),
             \rd,\textup{abs}\!\left(\!\frac{\xi^{\frac{1}{\alpha}}\rho_{\!D}^{\frac{1}{\alpha}}\dis}{\xi^{\frac{2}{\alpha}}-\rho_{\!D}^{\frac{2}{\alpha}}}\!\right)\right)}{\pi\rd^2(-1)^{\textbf{1}\!(\xi>\rho_{\!D})+1}}, \\ \quad\quad\quad\quad\quad\quad\quad\quad\quad\quad\quad\quad{\rdd\leq \dis<\rddd}; \\
             1, \quad\quad\quad\quad\quad\quad\quad\quad\quad\quad\quad\quad\quad\quad{\dis\geq \rddd};
             \end{cases}  %
  \end{align}
  \fi
\noindent where $\rdd=\textup{abs}\!\left(\!1\!-\!\left(\!\frac{\rho_{\!D}}{\xi}\!\right)^{\!\frac{1}{\alpha}}\!\right)\!\rd$, $\rddd=\left(\!1\!+\!\left(\!\frac{\rho_{\!D}}{\xi}\right)^{\!\frac{1}{\alpha}}\!\right)\!\rd$, $\xi$ is the mode selection threshold and $\psi(\cdot,\cdot,\cdot)$ is defined in~\eqref{eq:psi} in Lemma 1. For $\xi=\rho_{\!D}$, we have $p_{\DD}(\dis)=1-\frac{\rd^2\textup{acos}\!\left(\frac{\dis}{2\rd}\right)-\frac{\dis}{2}\sqrt{\rd^2-\frac{\dis^2}{4}}}{\pi\rd^2}$ when $\dis<2\rd$, while $p_{\DD}(\dis)=1$ if $\dis\geq2\rd$.

Under the different path-loss exponent scenario, this probability can be approximated by
\ifCLASSOPTIONonecolumn
      \begin{align}\label{eq:pactive_3}
p_{\DD}(\dis)\approx  1+\sum_{n=1}^{N}\left(-1\right)^{n}\binom {N}{n}\frac{2\dis^{2\frac{\alpha_{\!C}}{\alpha_{\!D}}}\left(nN\xi\right)^{\frac{2}{\alpha_{\!D}}}\Gamma\!\left[-\frac{2}{\alpha_{\!D}},\frac{\dis^{\alpha_{\!C}}nN\xi}{(N!)^{1/N}\rho_{\!D}\rd^{\alpha_{\!D}}}\right]}{\rd^2\alpha_{\!D}\left((N!)^{1/N}\rho_{\!D}\right)^{\frac{2}{\alpha_{\!D}}}},
  \end{align}
  \else
      \begin{align}\label{eq:pactive_3}
p_{\DD}(\dis)\approx & 1+\sum_{n=1}^{N}\left(-1\right)^{n}\binom {N}{n}\frac{2\dis^{2\frac{\alpha_{\!C}}{\alpha_{\!D}}}\left(nN\xi\right)^{\frac{2}{\alpha_{\!D}}}}{\rd^2\alpha_{\!D}\left((N!)^{1/N}\rho_{\!D}\right)^{\frac{2}{\alpha_{\!D}}}}\nonumber\\
&\times\Gamma\!\left[-\frac{2}{\alpha_{\!D}},\frac{\dis^{\alpha_{\!C}}nN\xi}{(N!)^{1/N}\rho_{\!D}\rd^{\alpha_{\!D}}}\right],
  \end{align}
  \fi
\noindent where $N$ is the parameter of a Gamma distribution which is used to formulate the approximation\footnote{By comparing with simulation results, we have verified that the average number of successful D2D transmissions obtained using this approximation is accurate when $N=6$.}.

  \noindent Proof: See Appendix~\ref{app_active2}.
  \end{prop}

\subsection{Spectrum reuse ratio}
Since we have employed the mode selection scheme, not all p-DUEs are in D2D mode. To evaluate the efficiency of our considered mode selection scheme, we propose a metric, namely the spectrum reuse ratio, which quantifies the average fraction of DUEs that can successfully transmit among all DUEs. For analytical tractability\footnote{Note that a more accurate metric is the average of the ratio $\frac{\textrm{number of successful D2D transmissions}}{\textrm{number of DUEs}}$. However, such a metric is very difficult to obtain. Instead, we consider the metric in~\eqref{eq:usespectrum}. It can be numerically verified that the values for these two metrics are very close to each other.}, spectrum reuse ratio is given by
\begin{align}\label{eq:usespectrum}
\tau&=\frac{\textrm{average number of successful D2D transmissions}}{\textrm{average number of DUEs}}\nonumber\\
&=\frac{\bar{M}}{\bar{M}_{\DD}},
\end{align}
\noindent where $\bar{M}$ is given in~\eqref{eq:spectral2}, and $\bar{M}_{\DD}$ is the average number of DUEs, which can be obtained as
\ifCLASSOPTIONonecolumn
\begin{align}\label{eq:averagenumber}
\bar{M}_{\DD}=&\mathbb{E}_{\totalphi,r_d}\left\{\sum\limits_{x\in\totalphi}\textbf{1}\!\left(\rho_{\!D} r_{d}^{\alpha_{\!D}}<\xi r_{c}^{\alpha_{\!C}}\right) \right\}=\int_0^{\rd}\!\!\!\left(\int_0^{\RM}\!\!\!\textbf{1}\!\left(\rho_{\!D} r_{d}^{\alpha_{\!D}}<\xi r_{c}^{\alpha_{\!C}}\right)2\pi r_c \lambda\,\textup{d}r_c\right)\! \frd\,\textup{d}r_d  \nonumber\\
=& \lambda\pi\!\int_{0}^{\tilde{\rd}}\!\!\!\int_{r_d^{\!\frac{\alpha_{\!D}}{\alpha_{\!C}}}\!\left(\!\frac{\rho_{\!D}}{\xi}\!\right)^{\!\frac{1}{\alpha_{\!C}}}}^{\RM}\!\!2r_c\frd\,\textup{d}r_c\,\textup{d}r_d\nonumber\\
=&\lambda\pi\RM^2\left(\frac{\tilde{\rd}^2}{\rd^2}-\frac{\alpha_{\!C}}{\alpha_{\!C}+\alpha_{\!D}}\frac{(\rho_{\!D}/\xi)^{\frac{2}{\alpha_{\!C}}} \tilde{\rd}^{2\frac{\alpha_{\!D}}{\alpha_{\!C}}+2}}{\RM^2\rd^2}\right).
\end{align}
\else
\begin{align}\label{eq:averagenumber}
&\bar{M}_{\DD}=\mathbb{E}_{\totalphi,r_d}\left\{\sum\limits_{x\in\totalphi}\textbf{1}\!\left(\rho_{\!D} r_{d}^{\alpha_{\!D}}<\xi r_{c}^{\alpha_{\!C}}\right) \right\}\nonumber\\
&=\int_0^{\rd}\!\!\!\left(\int_0^{\RM}\!\!\!\textbf{1}\!\left(\rho_{\!D} r_{d}^{\alpha_{\!D}}<\xi r_{c}^{\alpha_{\!C}}\right)2\pi r_c \lambda\,\textup{d}r_c\right)\! \frd\,\textup{d}r_d  \nonumber\\
&= \lambda\pi\!\int_{0}^{\tilde{\rd}}\!\!\!\int_{r_d^{\!\frac{\alpha_{\!D}}{\alpha_{\!C}}}\!\left(\!\frac{\rho_{\!D}}{\xi}\!\right)^{\!\frac{1}{\alpha_{\!C}}}}^{\RM}\!\!2r_c\frd\,\textup{d}r_c\,\textup{d}r_d\nonumber\\
&=\lambda\pi\RM^2\!\!\left(\!\frac{\tilde{\rd}^2}{\rd^2}\!-\!\frac{\alpha_{\!C}}{\alpha_{\!C}+\alpha_{\!D}}\frac{(\rho_{\!D}/\xi)^{\frac{2}{\alpha_{\!C}}} \tilde{\rd}^{2\frac{\alpha_{\!D}}{\!\alpha_{\!C}}+2}}{\RM^2\rd^2}\!\right).
\end{align}
\fi

\subsection{Summary}
Summarizing, for the underlay in-band D2D communication with the considered mode selection scheme in a disk-shaped cellular network region, following the system model in Section~\ref{sec:systemmodel}, we can calculate:
 \begin{enumerate}[(i)]
   \item outage probability at the BS by combining~\eqref{eq:MGFTotalbs} and~\eqref{eq:MGF_BS_lt_final}~\redcom{in Proposition~\ref{prop_schemeBS}} and substituting into~\eqref{eq:outage_naka};
   \item conditional outage probability at a DRx by combining~\eqref{eq:MGFTotalDRx},~\eqref{eq:MGF_DRx_lt_final} \redcom{in Proposition~\ref{prop_scheme2}} and~\eqref{eq:MGF_DRx_CUE_final} \redcom{in Corollary~\ref{prop_CUE}} and substituting into~\eqref{eq:outage_naka};
   \item average number of successful D2D transmissions by substituting the conditional outage probability at a DRx,~\eqref{eq:densityfunction} in \redcom{Proposition~\ref{theorem_density}} and \eqref{eq:pactive_2} (or~\eqref{eq:pactive_3}) \redcom{in Proposition~\ref{prop_active}}, into~\eqref{eq:spectral2} \redcom{in Proposition~\ref{prop:link}};
   \item spectrum reuse ratio by finding the ratio of average number of successful D2D transmissions and $\bar{M}_{\DD}$ in~\eqref{eq:averagenumber}.
 \end{enumerate}
Note that the evaluation of the analytical results requires the differentiation and integration of the MGFs, which can be easily implemented using mathematical packages such as Mathematica.


\section{Results}\label{sec:result}
In this section, we present the numerical results to study the impact of the D2D system parameters (i.e., the p-DUE's node density $\lambda$ and the receiver sensitivity of DRx $\rho_{\!D}$) on the outage probability, the average number of successful D2D transmissions and spectrum reuse ratio. To validate our derived results, the simulation results are generated using MATLAB, which are averaged over $10^6$ simulation runs. Note that in the simulations, all DRxs are confined in the region $\area$. Unless specified otherwise, the values of the main system parameters shown in Table~\ref{tb:3} are used. We assume a cell region radius of $\RM = 500$ m. \redcom{The vast majority of the D2D literature has considered either $\alpha_C=\alpha_D$ (i.e.,~\cite{ElSawy-2014,Lin-2014,Marshall-2015,Stefanatos-2015,Yu-2011,Lee-2015,Ni-2015,7236879}) or $\alpha_C$ is slightly smaller than $\alpha_D$ (i.e.,~\cite{Yu-2014,George-2015,Ali-2014,7005537,7137678}).} Hence we adopt the following path-loss exponent values when generating the main results: \redcom{$\alpha_C = 3.5, 3.75, 4$ and $\alpha_D = 4$}.\footnote{\redcom{Consideration of multi-slope model~\cite{7061455} is outside the scope of this paper}.}
\begin{table}
\centering
\caption{Main System Parameter Values.}
\label{tb:3}
\begin{tabular}{|l|l|c|}\hline
Parameter & Symbol & Value \\
\hline
p-DUE's node density & $\lambda$ & $5*10^{-5}$ users/m$^2$\\ \hline
p-DUE's transmission range & $\rd$ & $35$ m\\ \hline
Receiver sensitivity for BS& $\rho_{\BS}$ & $-80$ dBm \\ \hline
Receiver sensitivity for DRx& $\rho_{\!D}$ & $-70$ dBm \\ \hline
SIR threshold & $\gamma$ & $0$ dB \\ \hline
\end{tabular}
\vspace{-0.1 in}
\end{table}

\subsection{Model validation}
\begin{figure}
\centering
\subfigure[Outage probability at BS, $P_{\out}^{\BS}(\gamma)$.]{\label{outage_BS}\includegraphics[width=0.47\textwidth]{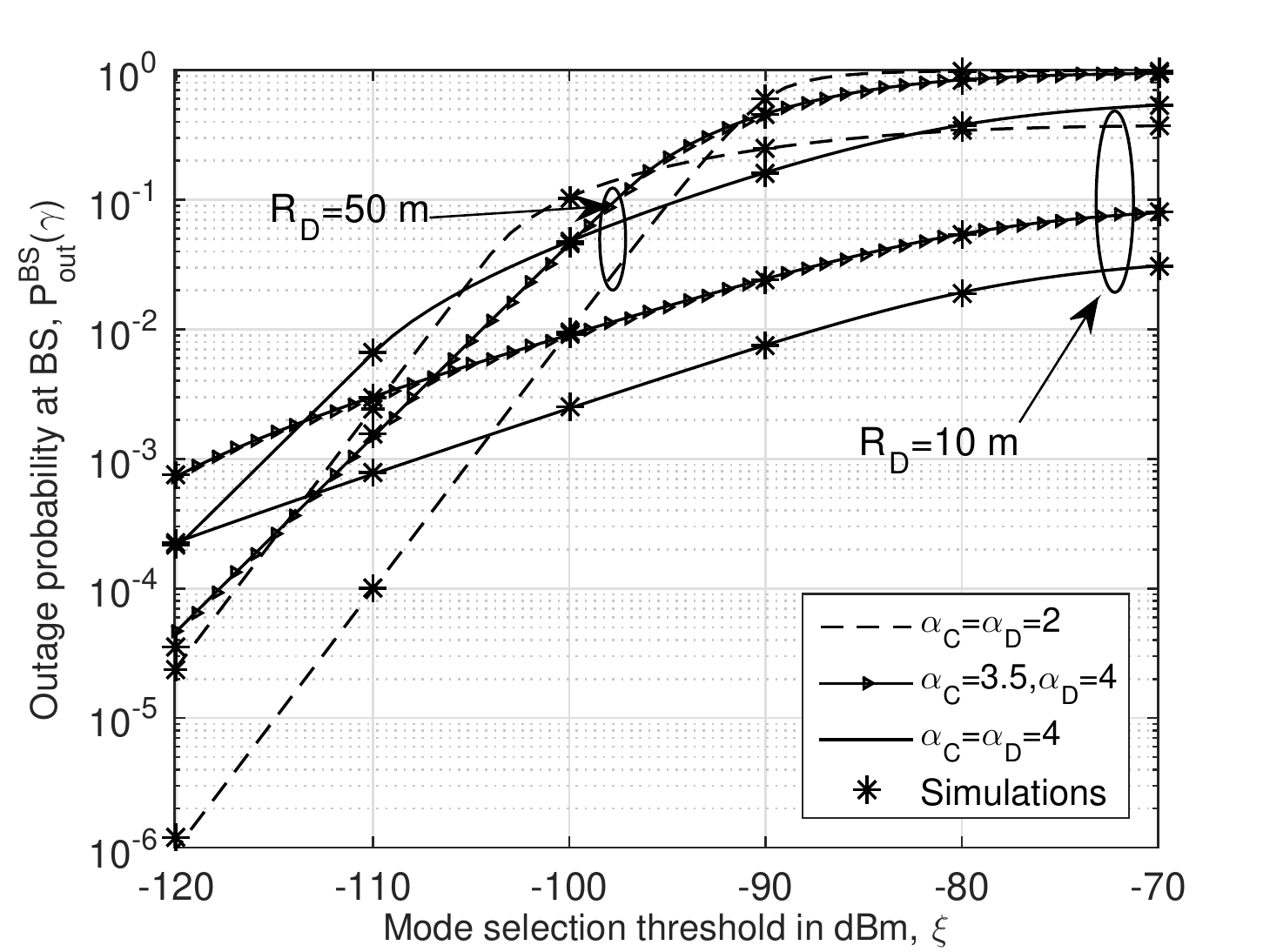}}
\subfigure[Average number of successful D2D transmissions, $\bar{M}$.]{ \label{thru}\includegraphics[width=0.47\textwidth]{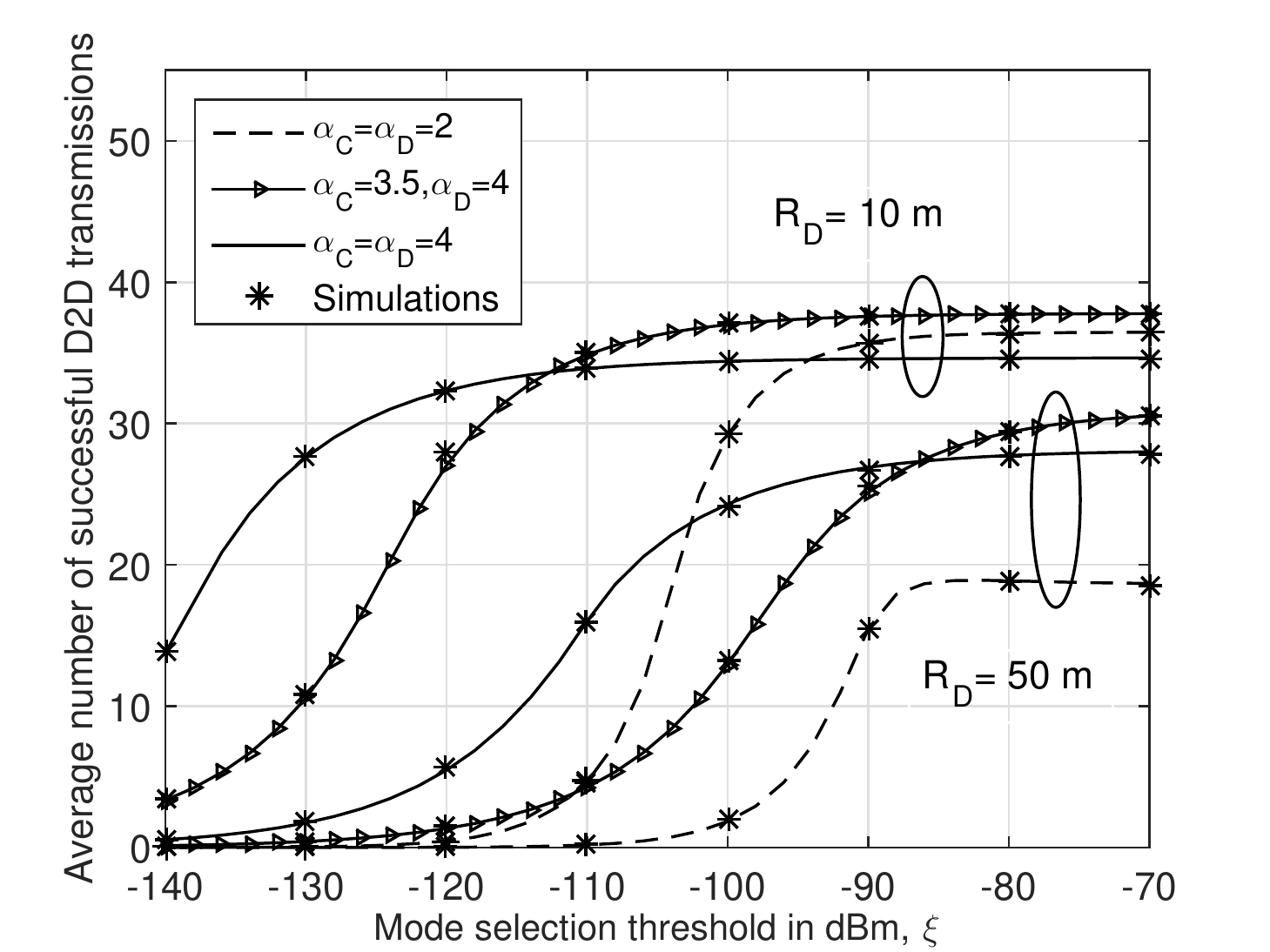}}
\vspace{-0.05 in}
\caption{\redcom{Outage probability at BS and average number of successful D2D transmissions versus the mode selection threshold $\xi$ for $\rd=10$ m and $\rd=50$ m, respectively.}} \label{fig:outageBSandlink}
\vspace{-0.1 in}
\end{figure}
In this subsection, we illustrate the accuracy of our derived results. Fig.~\ref{fig:outageBSandlink} plots the outage probability at BS and the average number of successful D2D transmissions versus the mode selection threshold $\xi$ for different path-loss exponent sets, for $\rd=10$ m and $\rd=50$ m, respectively. The fading on the desired cellular link and the desired D2D link are assumed to be Rayleigh fading and Nakagami fading with $m=3$, respectively. The analytical curves in Fig.~\ref{outage_BS} are plotted using Proposition~\ref{prop_schemeBS}, i.e., substituting~\eqref{eq:MGFTotalbs} and~\eqref{eq:MGF_BS_lt_final} into~\eqref{eq:outage_naka}, while the curves in Fig.~\ref{thru} are plotted using the combination of Propositions~2-5, and Corollary~\ref{prop_CUE}. From both figures, we can see that the analytical results match closely with the simulation results even when the mode selection threshold is small (i.e., probability of being DUE is small) or the radius of the p-DUE's transmission range is relatively large (i.e., 10$\%$ of the cell radius). This confirms the accuracy of our derived approximation results. In addition, as shown in Fig.~\ref{fig:outageBSandlink}, both the outage probability at the BS and the average number of successful D2D transmissions increase as the  mode selection threshold increases. This is because as mode selection threshold increases, more p-DUEs are allowed to be in underlay D2D mode which improves the average number of successful D2D transmissions. However, the increase in mode selection threshold degrades the outage performance at the BS since more interferers are involved.
\subsection{Outage probability at DRx: Location-dependent performance}
\begin{figure}[t]
        \centering
        \includegraphics[width=0.47  \textwidth]{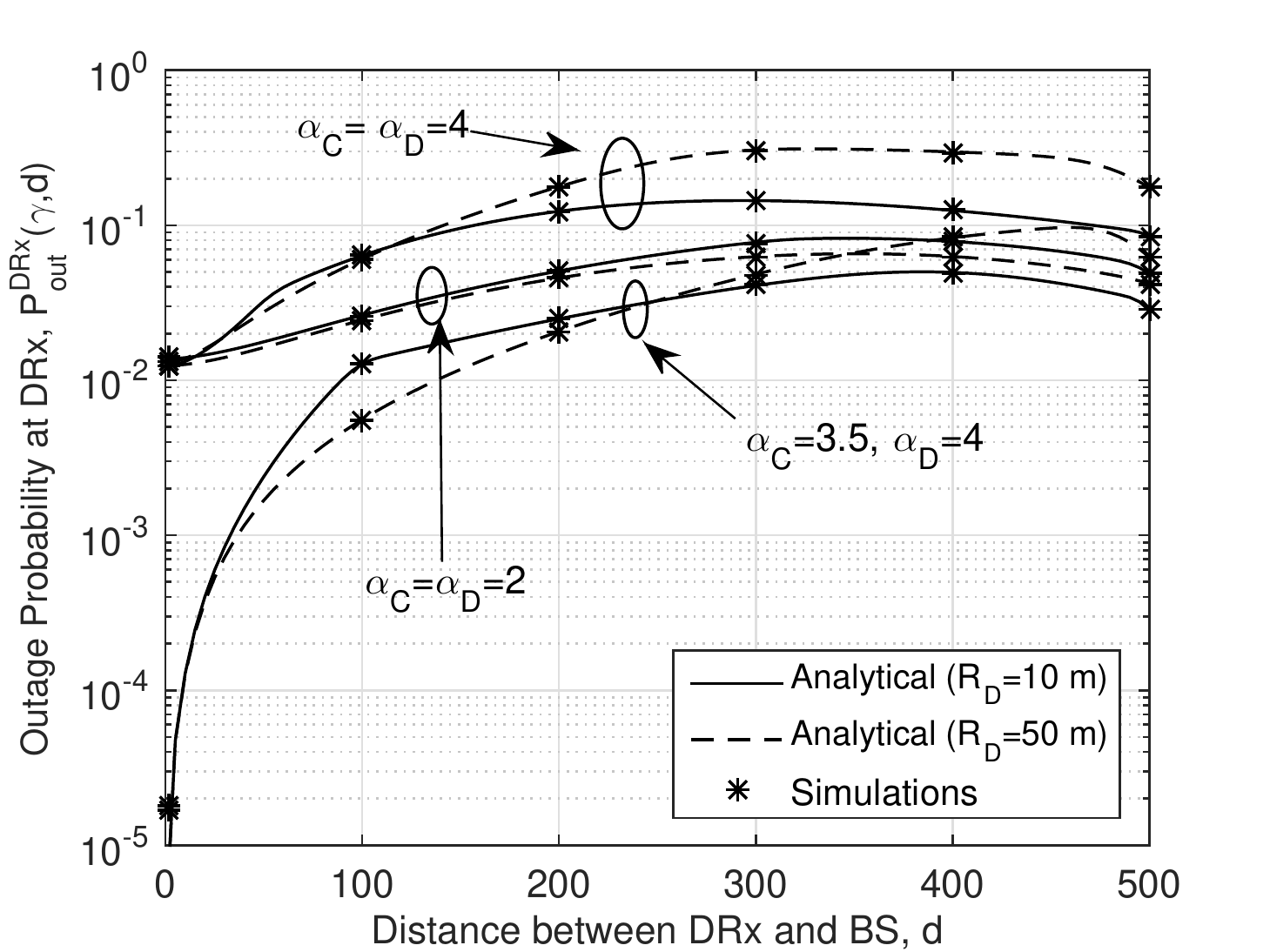}
               \vspace{-0.05 in}
        \caption{\redcom{Outage probability at DRx, $P_{\out}^{\DRx}(\gamma,d)$, versus the distance between BS and DRx $d$ for $\rd=10$ m and $\rd=50$ m, respectively.}}
        \label{outage_d}
        \vspace{-0.05 in}
\end{figure}

Fig.~\ref{outage_d} plots the outage probability at a typical DRx versus its distance to the BS with different path-loss exponent sets, for $\rd=10$ m and $\rd=50$ m, respectively. The simulation results are also presented and match well with the analytical results, which again validates our analytical results. As illustrated in Fig.~\ref{outage_d}, the outage probability at the DRx varies greatly with the DRx location, which highlights the importance of characterizing the location-dependent performance. The general trends are that the outage probability firstly increases as the distance between DRx and BS increases and then decreases when the DRx is close to the cell-edge. These trends can be explained as follows. When the DRx is close to the BS, there are fewer number of p-DUEs that are in underlay D2D mode due to the mode selection scheme. Thus, interference is less and the outage probability is low. As the DRx gradually moves away from the BS, more interfering nodes are present and the outage probability increases. However, once the DRx is close to the cell-edge, the number of interfering DUEs decreases due to the boundary effect, and the outage probability decreases.

\subsection{Effects of D2D user's density}\label{results:d}
In this subsection, we investigate the effect of p-DUE's node density $\lambda$ on the average number of successful D2D transmissions and spectrum reuse ratio (i.e., the average fraction of DUEs that can successfully transmit among all DUEs). Since both the outage probability at the BS and the average number of successful D2D transmissions are increasing functions of the mode selection threshold $\xi$, as shown in Fig.~\ref{fig:outageBSandlink}, we have adopted the following method to investigate the effects of D2D user's density:
\begin{itemize}
  \item Given a QoS at the BS, for each p-DUE's node density $\lambda$, using~\eqref{eq:outage_naka},~\eqref{eq:MGFTotalbs} and~\eqref{eq:MGF_BS_lt_final}, we can find the mode selection threshold $\xi$ satisfying the QoS at the BS;
  \item Using the mode selection threshold $\xi$ that satisfies the QoS at BS, the average number of successful D2D transmissions $\bar{M}$ can be calculated for each $\lambda$. This obtained $\bar{M}$ value can be regarded as the maximum average number of successful underlay D2D transmission achieved by the system. We can then work out the corresponding spectrum reuse ratio.
\end{itemize}
\begin{figure*}[!t]
\centering
\subfigure[$\rho_{D}=-70$ dBm.]{\label{link_M1}\includegraphics[width=0.4\textwidth]{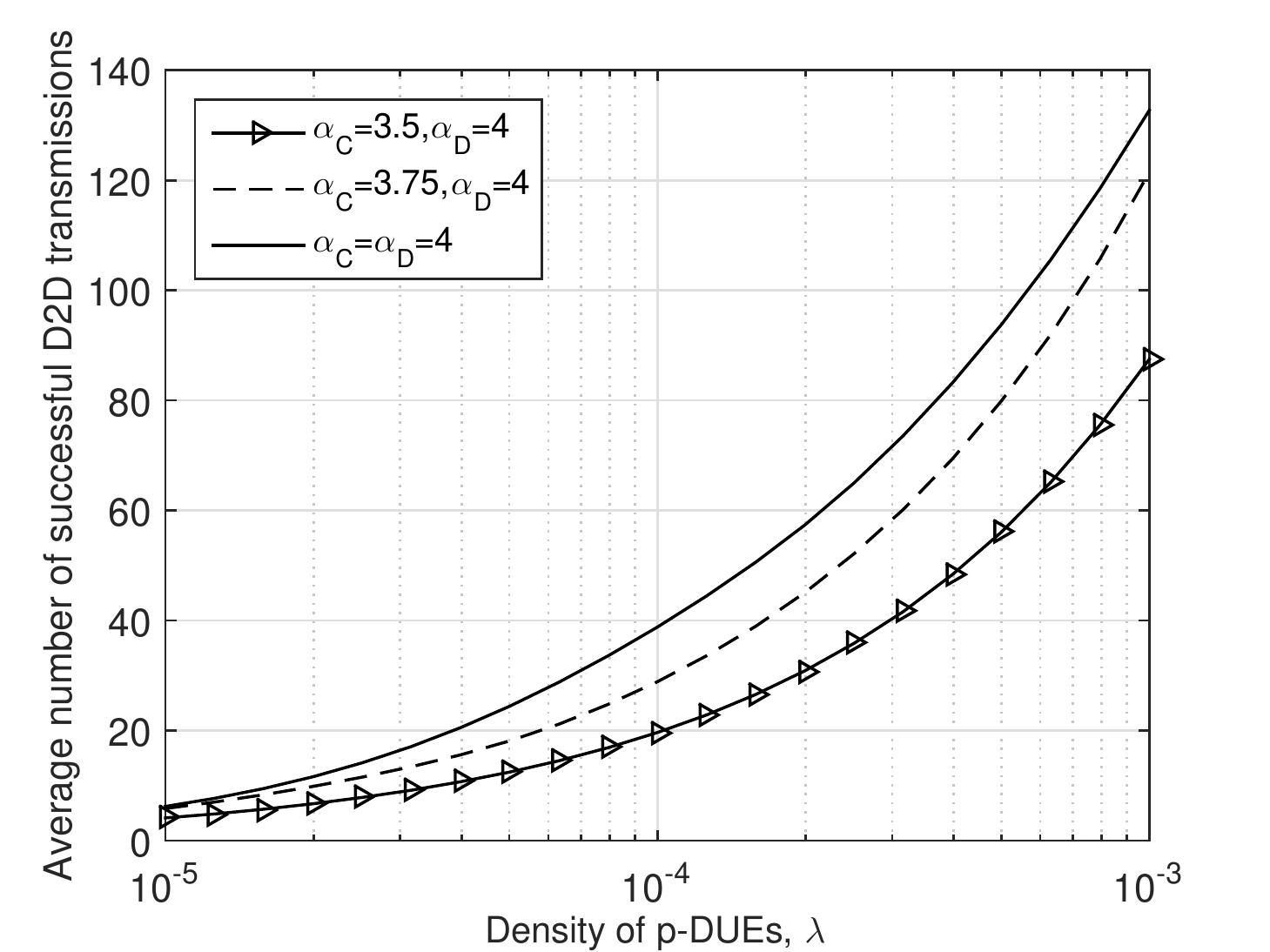}}
\subfigure[$\rho_{D}=-90$ dBm.]{\label{link_M2}\includegraphics[width=0.4\textwidth]{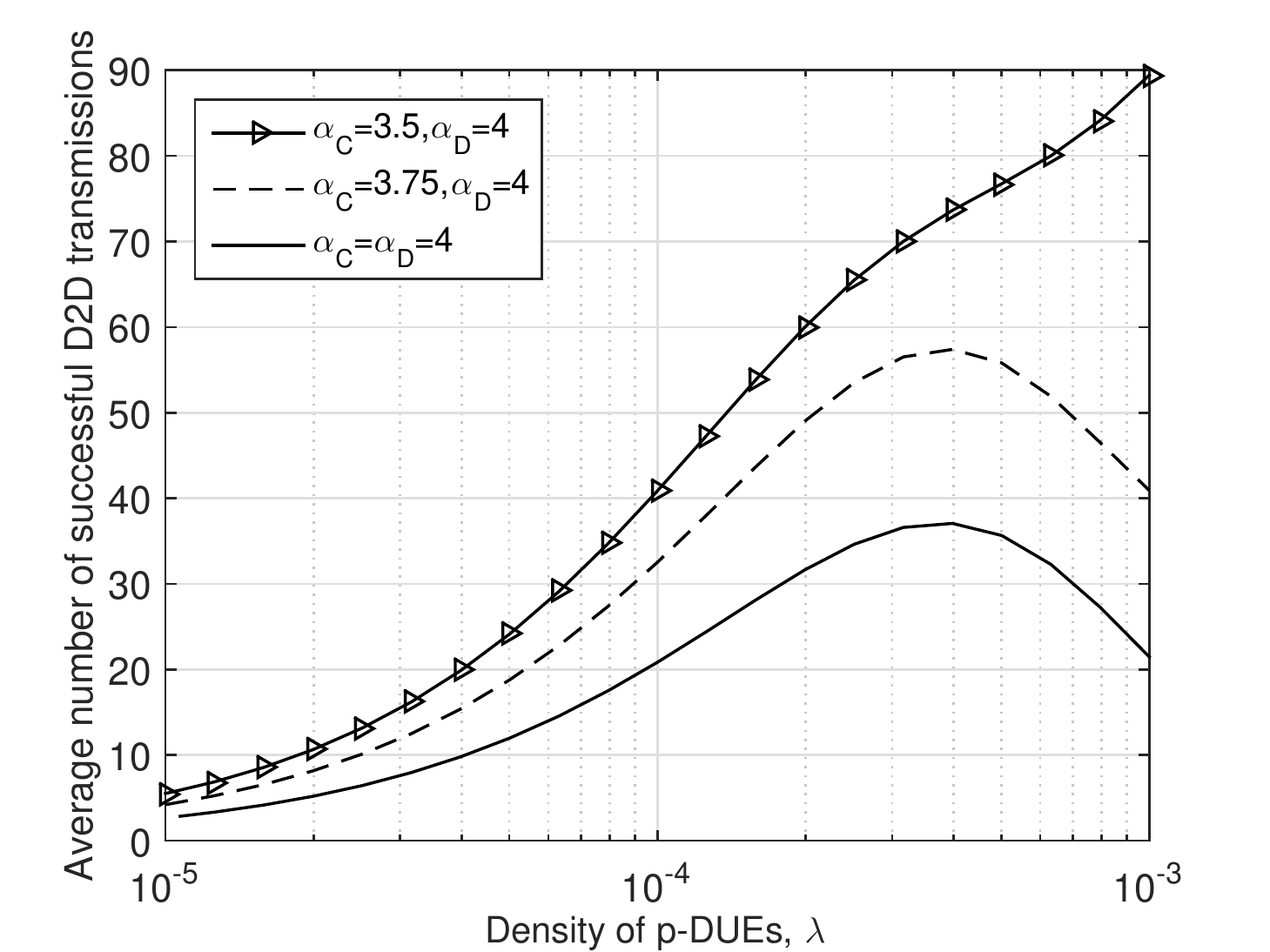}}
\subfigure[$\rho_{D}=-70$ dBm.]{\label{ratio_M1}\includegraphics[width=0.4\textwidth]{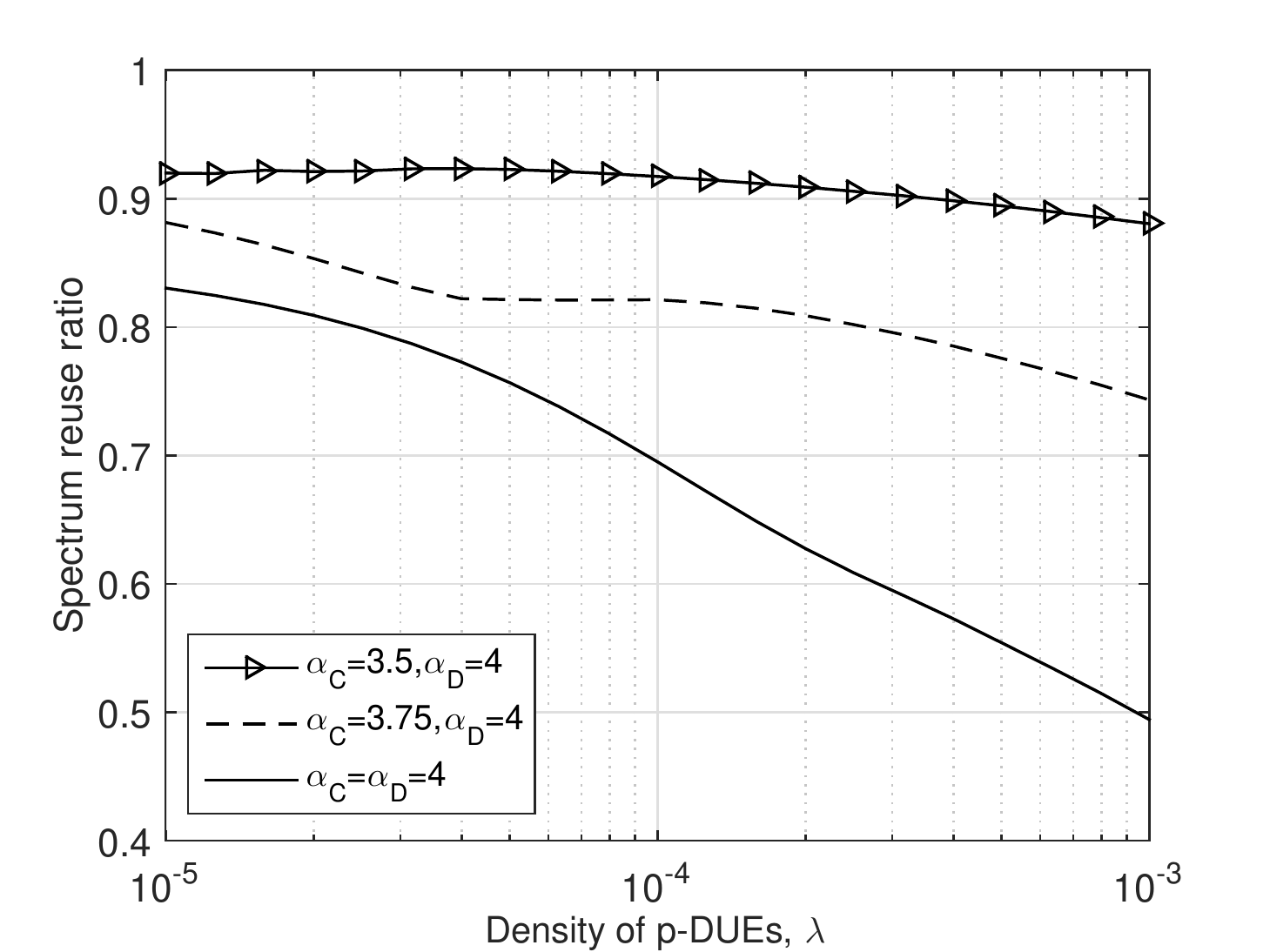}}
\subfigure[$\rho_{D}=-90$ dBm.]{\label{ratio_M2}\includegraphics[width=0.4\textwidth]{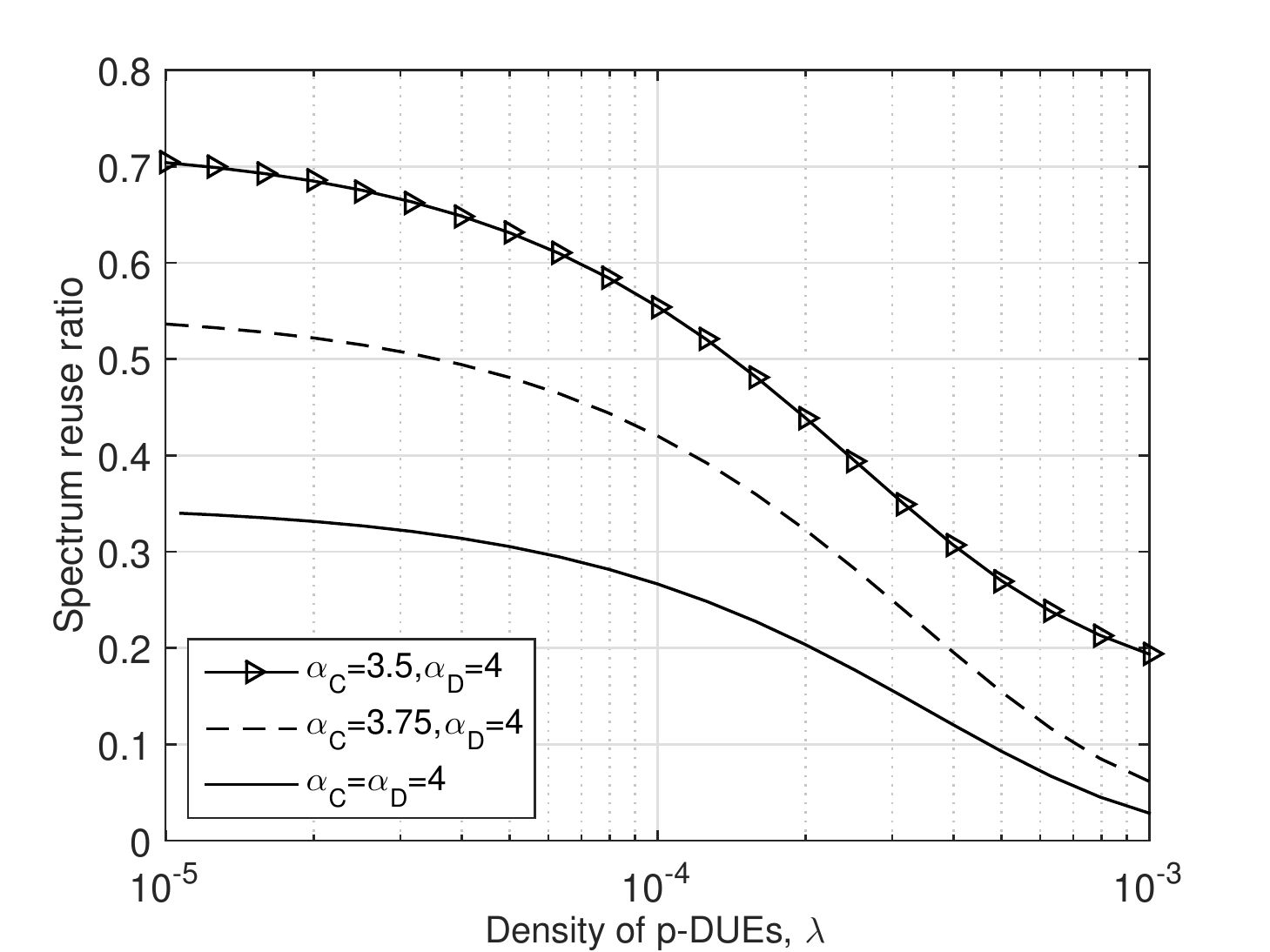}}
\caption{\redcom{Average number of successful D2D transmissions $\bar{M}$ and spectrum reuse ratio $\tau$ versus the node density of p-DUEs $\lambda$, with different receiver sensitivity of DRx $\rho_{\!D}$, and QoS constraint $P_{\out}^{\BS}(\gamma)=10^{-2}$.}}
        \label{link_M}
              \vspace{-0.05 in}
  \end{figure*}

Fig.~\ref{link_M} plots the average number of successful D2D transmissions and spectrum reuse ratio versus the node density of p-DUEs for QoS constraint at the BS $P_{\out}^{\BS}(\gamma)=10^{-2}$ and different DRx's receiver sensitivity. We assume the fading on all the links to be Rayleigh fading. From Figs.~\ref{link_M1} and~\ref{ratio_M1}, we can see that the average number of successful D2D transmissions increases with increasing node density of p-DUE, however the spectrum reuse ratio decreases. This trend can be explained as follows. When the node density is higher, the probability of being in D2D mode is reduced to maintain the QoS at the BS. However, the overall node density is large which means that the number of DUEs is still large. Thus, the average number of successful D2D transmissions, which is mainly affected by the number of DUEs under this scenario, increases when the node density of p-DUEs increases. In contrast, lesser number of DUEs are likely to transmit successfully when the number of interfering DUEs is large, which leads to the decreasing trend of spectrum reuse ratio.

From Fig.~\ref{link_M2}, we can see that when the receiver sensitivity of DRx is smaller than that of BS, increasing p-DUE's node density beyond a certain limit can degrade the average number of successful D2D transmissions, especially when $\alpha_{\!C}$ and $\alpha_{\!D}$ \redcom{have very similar values}. This is due to the fact that the average number of successful D2D transmissions is determined by the number of DUEs and the outage probability at DRx. When $\rho_{\!D}$ is small, since there is a greater number of interfering DUEs nearby and the interference from CUE can be also severe \redcom{when $\alpha_{\!C}$ and $\alpha_{\!D}$ have very similar values}, the outage probability at DRx is high. Thus, $\bar{M}$ first increases and then decreases.

\redcom{From Fig.~\ref{ratio_M1}, we can see that if $\alpha_{\!C}$ is slightly smaller than $\alpha_{\!D}$ and $\rho_D$ is greater than $\rho_C$,} the decreasing trend for spectrum reuse ratio is almost negligible. In other words, the spectrum reuse ratio can be regarded as almost a constant and it does not degrade with increasing node density of p-DUE. Under such a case, increasing the p-DUE's node density is beneficial for underlay D2D communication.

\subsection{Effects of D2D user's receiver sensitivity}
\begin{figure*}[!t]
\centering
\subfigure[$\rho_{\BS}=-80$ dBm.]{\label{link_rhod1}\includegraphics[width=0.4\textwidth]{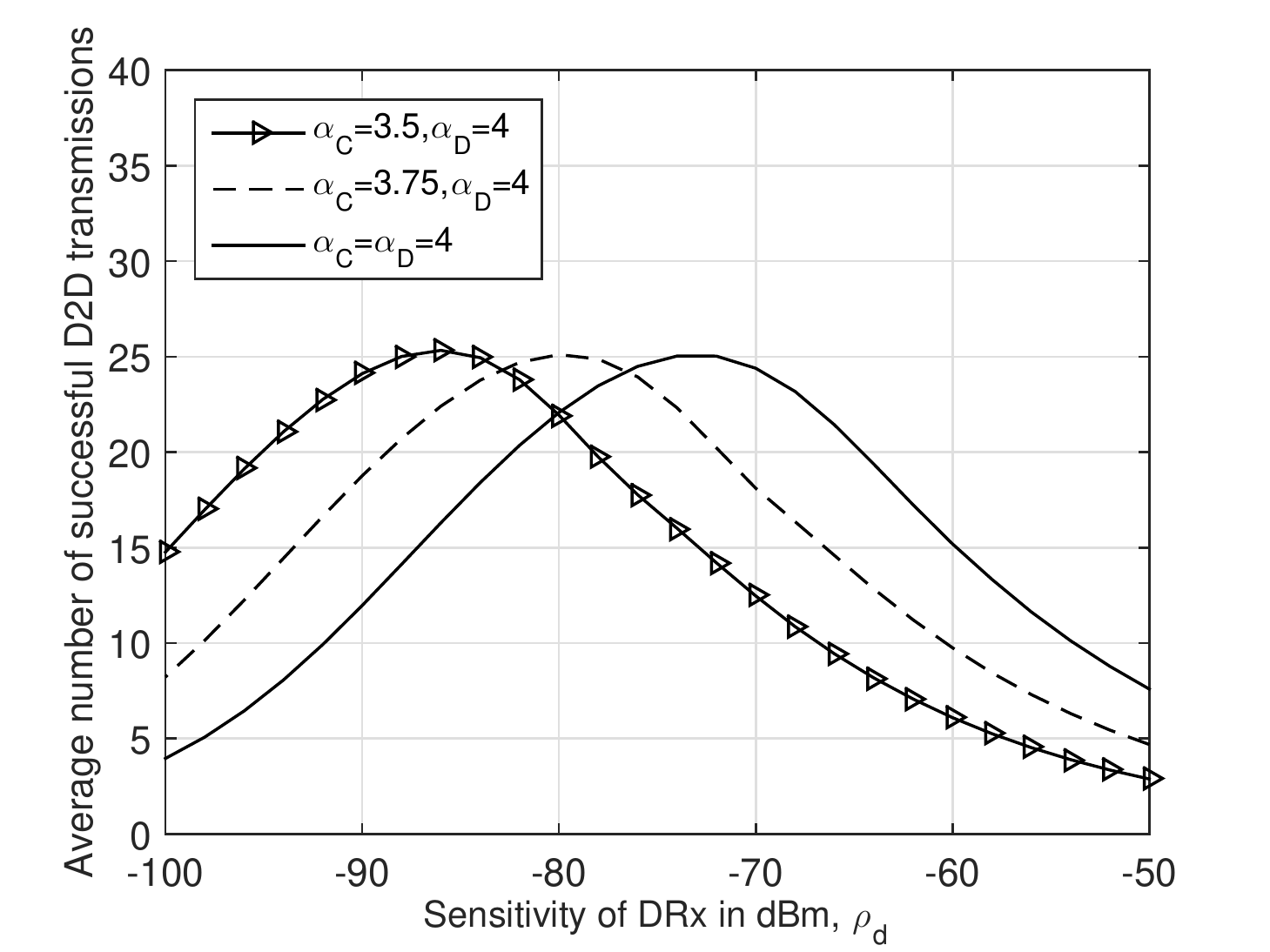}}
\subfigure[$\rho_{\BS}=-60$ dBm.]{ \label{link_rhod2}\includegraphics[width=0.4\textwidth]{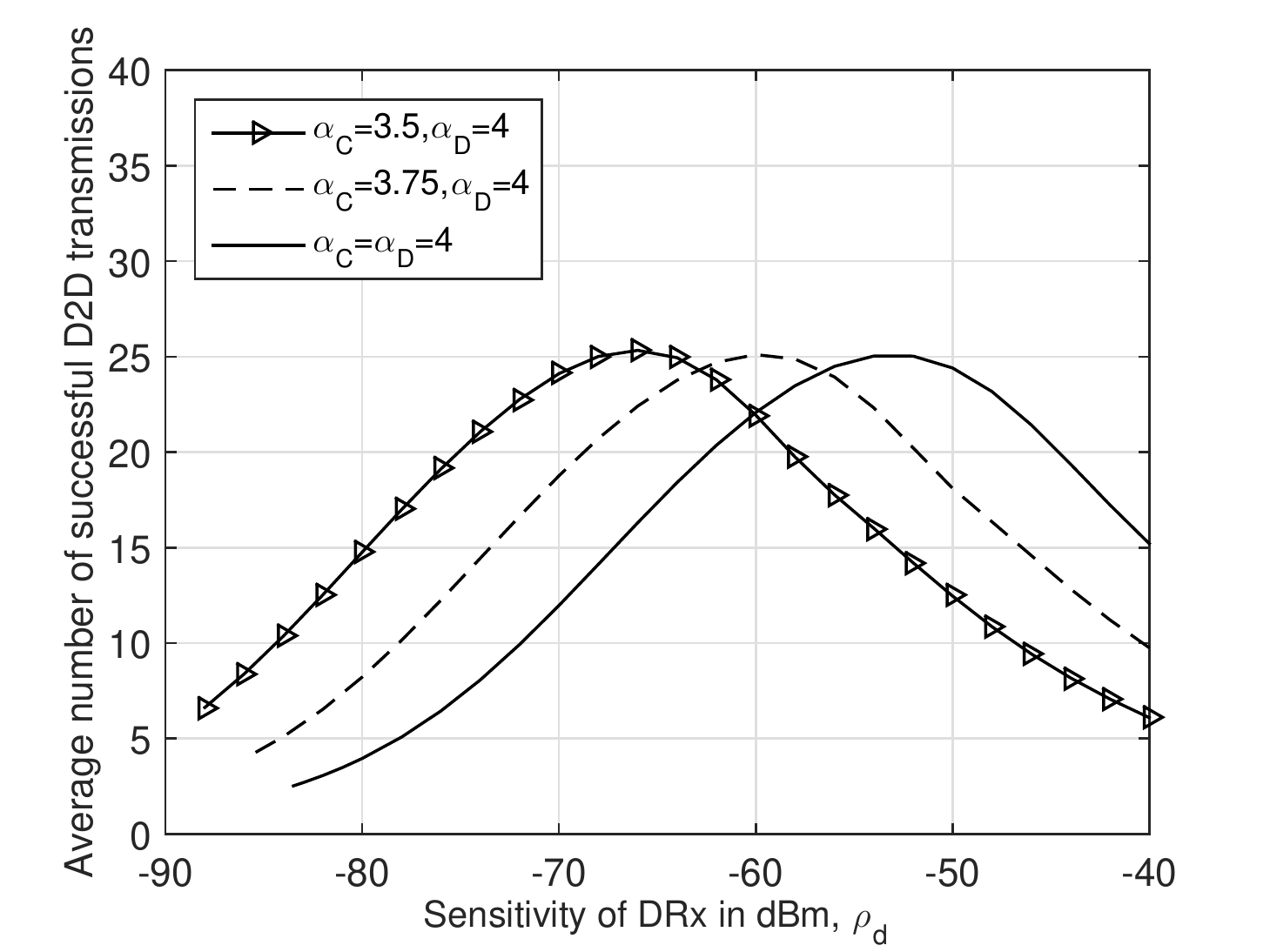}}
\subfigure[$\rho_{\BS}=-80$ dBm.]{\label{ratio_rhod1}\includegraphics[width=0.4\textwidth]{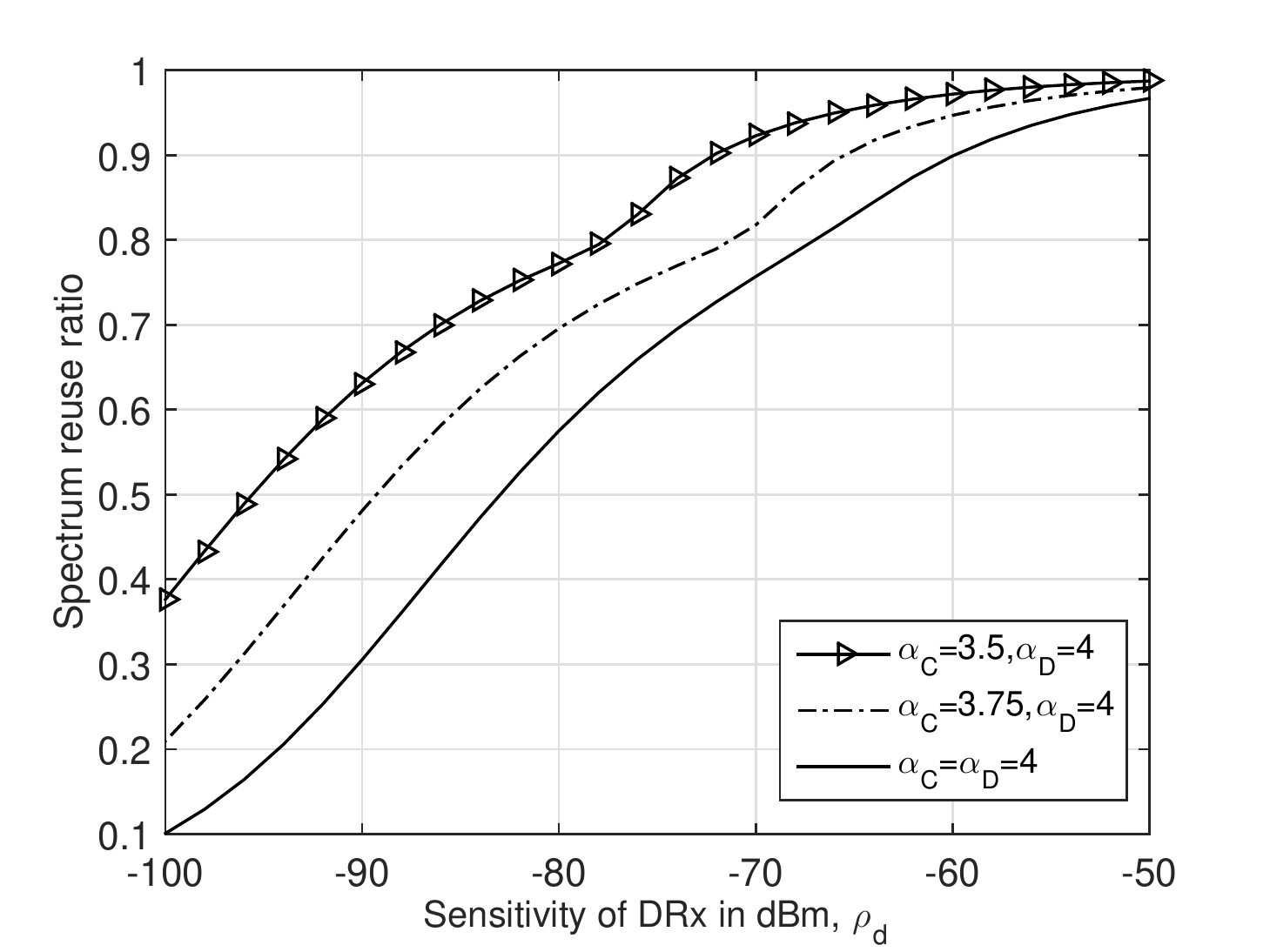}}
\subfigure[$\rho_{\BS}=-60$ dBm.]{ \label{ratio_rhod2}\includegraphics[width=0.4\textwidth]{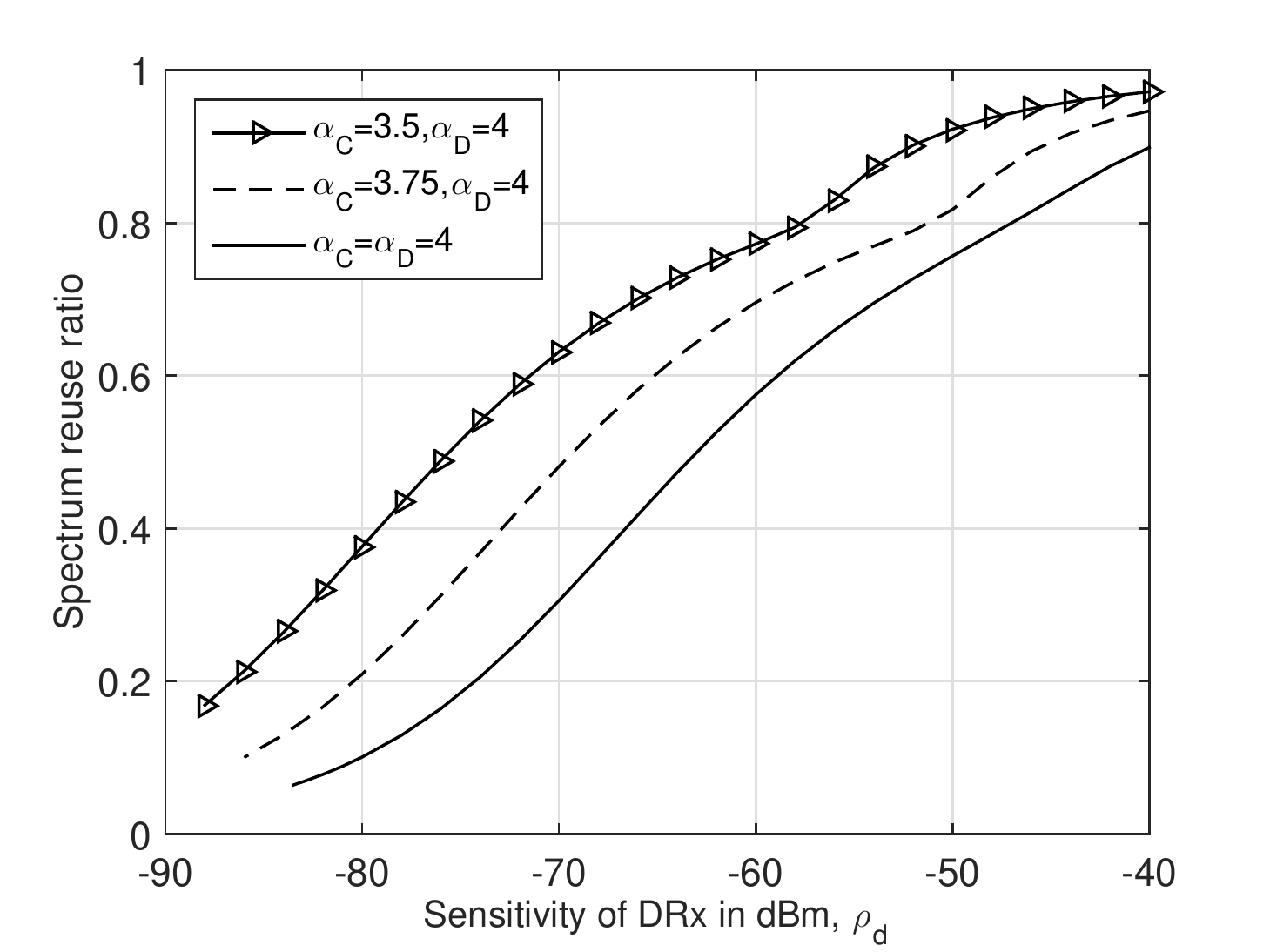}}
\caption{\redcom{Average number of successful D2D transmissions $\bar{M}$ and spectrum reuse ratio $\tau$ versus the DRx's receiver sensitivity $\rho_{\!D}$, with different receiver sensitivity of BS $\rho_{\BS}$, and QoS constraint $P_{\out}^{\BS}(\gamma)=10^{-2}$.}}
        \label{link_rhod}
        \vspace{-0.05 in}
  \end{figure*}

In this subsection we examine the effect of DRx's receiver sensitivity $\rho_{\!D}$ on the average number of successful D2D transmissions and spectrum reuse ratio, adopting the same approach as explained in Section~\ref{results:d}. Fig.~\ref{link_rhod} plots the average number of successful D2D transmissions and related spectrum reuse ratio versus the DRx's receiver sensitivity  with QoS constraint at the BS $P_{\out}^{\BS}(\gamma)=10^{-2}$, for different path-loss exponent sets and receiver sensitivity of BS $\rho_{\BS}$.

From Figs.~\ref{link_rhod1} and~\ref{link_rhod2}, we can see that, in general, as the receiver sensitivity decreases the average number of successful D2D transmissions increases at first and then decreases. These trends can be explained as follows. The average number of successful D2D transmissions is impacted by both the number of DUEs and the outage probability at DRxs. When $\rho_{\!D}$ is small, more p-DUEs are operating in D2D mode because their transmit power is small, so less interference is generated to the BS. Similarly, for the outage probability at DRx, although the total number of DUEs is large, the interference from surrounding DUEs is not severe due to the small receiver sensitivity. If we ignore the interference from CUE, the number of DUEs governs the network performance and the average number of successful D2D transmissions increases as the receiver sensitivity decreases. However, we cannot ignore the interference from CUE, especially when the value of $\alpha_{\!C}$ is large (i.e., close to the value of $\alpha_{\!D}$). Under such a scenario, the transmit power for CUE is large. Moreover, due to the smaller receiver sensitivity at DRxs, DRxs are more likely to be in outage. Consequently, the interplay of the number of DUEs and the outage probability at the DRx causes the average number of successful D2D transmissions to first increase and then decrease as the receiver sensitivity decreases.

Figs.~\ref{link_rhod1} and~\ref{link_rhod2} also show that for different path-loss exponent sets ($\alpha_{\!C}$ and $\alpha_{\!D}$), the maximum value of the average number of successful D2D transmissions occurs at different receiver sensitivity values. For example, \redcom{if $\alpha_{\!C}=\alpha_{\!D}$}, $\bar{M}$ reaches its maximum value when the value of $\rho_{\!D}$ is greater than $\rho_{\!C}$. However, \redcom{if $\alpha_{\!C}$ is smaller than $\alpha_{\!D}$,} then a smaller receiver sensitivity of DRx results in the maximum $\bar{M}$. That is to say, as the value of $\alpha_{\!C}$ decreases, the required receiver sensitivity of DRx to achieve the maximum average number of successful D2D transmissions becomes smaller. \redcom{Note that when $\rho_{\!C}$ is far greater than $\rho_{\!D}$, although all p-DUEs are in D2D mode, the outage probability at the BS will still be lower than $10^{-2}$. Hence, $\bar{M}$ cannot be computed and the curves are incomplete in Figs.~\ref{link_rhod2} and~\ref{ratio_rhod2} for certain cases.}

Figs.~\ref{ratio_rhod1} and~\ref{ratio_rhod2} show that the spectrum reuse ratio generally decreases as the DRx's receiver sensitivity decreases. Additionally, when the path-loss exponent on the \redcom{cellular link is slightly lower than the path-loss exponent on the D2D link}, then the decreasing amount in the spectrum reuse ratio is less for the different cases considered.

\section{Conclusion}\label{sec:conclusion}
In this paper, we proposed a framework to analyze the performance of underlay in-band D2D communication inside a finite cellular region. We adopted a mode selection scheme for potential D2D users to manage the intra-cell interference experienced by the BS. Using stochastic geometry, we derived approximate yet accurate analytical results for the outage probability at the BS and a typical DRx, the average number of successful D2D transmissions and spectrum reuse ratio.

Our derived results showed that the outage probability relies strongly on the location of DRx. They also allowed the impact of the D2D system parameters on both the average number of successful D2D transmissions and spectrum reuse ratio to be determined. For example, it is observed that, given the QoS constraint at the BS, as the D2D users' node density increases, the spectrum reuse ratio decreases. \redcom{When the receiver sensitivity of the DRx is greater than the receiver sensitivity of the BS, the average number of successful D2D transmissions increases. Moreover, when the path-loss exponent on the cellular link is slightly lower than that on the D2D link, the decreasing trend for spectrum reuse ratio can become negligible.} This indicates that an increasing level of D2D communication can be beneficial in future networks and provides design guidelines in the practical communication systems with D2D communication. Future work can analyze the impact of imperfect inter-cell interference cancellation for D2D communication in a finite multi-cell scenario.

\appendices

\section{Derivation of Equation~\eqref{eq:outage_naka}: Outage Probability }\label{app_outage}
\begin{proof}
Rearranging~\eqref{eq:outage}, we have the outage probability as
\ifCLASSOPTIONonecolumn
\begin{align}\label{eq:outage_de1}
P_{\out}^{\kappa}(\gamma)=&\mathbb{E}_{\Inftagg^{\kappa},g_0}\left\{\Pr\left(g_0<\frac{\gamma}{\rho}\Inftagg^{\kappa}\right)\right\}
=\mathbb{E}_{\Inftagg^{\kappa}}\left\{F_{g_0}\left(\frac{\gamma}{\rho}\Inftagg^{\kappa}\right)\right\},
\end{align}
\else
\begin{align}\label{eq:outage_de1}
P_{\out}^{\kappa}(\gamma)=&\mathbb{E}_{\Inftagg^{\kappa},g_0}\!\left\{\!\Pr\!\left(g_0<\frac{\gamma}{\rho}\Inftagg^{\kappa}\right)\!\right\}
=\mathbb{E}_{\Inftagg^{\kappa}}\!\left\{\!F_{g_0}\!\left(\frac{\gamma}{\rho}\Inftagg^{\kappa}\right)\!\right\},
\end{align}
\fi
\noindent where $F_{g_0}(\cdot)$ is the cumulative distribution function (CDF) of the fading power gain on the reference link. Since we assume Nakagami-m fading with integer $m$ for the reference link, $g_0$ follows the Gamma distribution with mean 1 and shape parameter $m$, and its CDF is given by $F_{g_0}(x)=1-\sum_{t=0}^{m-1}\frac{1}{t!}(mx)^{t}\exp(-mx)$. Hence, we can re-write~\eqref{eq:outage_de1} as
\begin{align}\label{eq:outage_de2}
P_{\out}^{\kappa}(\gamma)=&\mathbb{E}_{\Inftagg^{\kappa}}\left\{1-\sum_{t=0}^{m-1}\frac{1}{t!}\left(m\frac{\gamma}{\rho}\Inftagg^{\kappa}\right)^t\exp\left(-m\frac{\gamma}{\rho}\Inftagg^{\kappa}\right)\right\} \nonumber\\
=&1-\sum_{t=0}^{m-1}\frac{1}{t!}\mathbb{E}_{\Inftagg^{\kappa}}\left\{\left(m\frac{\gamma}{\rho}\Inftagg^{\kappa}\right)^t\exp\left(-m\frac{\gamma}{\rho}\Inftagg^{\kappa}\right)\right\}.
\end{align}

Note the MGF of $\Inftagg^{\kappa}$ is $\mathcal{M}_{\Inftagg^{\kappa}}(s)=\mathbb{E}_{\Inftagg^{\kappa}}\left\{\exp\left(-s\Inftagg^{\kappa}\right)\right\}$ and its $t$-th derivative with respect to $s$ is
$\frac{d^t}{d s^t}\mathcal{M}_{\Inftagg^{\kappa}}(s)=\mathbb{E}_{\Inftagg^{\kappa}}\left\{\frac{d^t\exp\left(-s\Inftagg^{\kappa}\right)}{d s^t}\right\}=\mathbb{E}_{\Inftagg^{\kappa}}\left\{\left(-\Inftagg^{\kappa}\right)^t\exp\left(-s\Inftagg^{\kappa}\right)\right\}$. By substituting $s=m\frac{\gamma}{\rho}$, we have
\begin{align}\label{eq:outage_de3}
\left.\frac{d^t}{d s^t}\mathcal{M}_{\Inftagg^{\kappa}}(s)\right|_{s=m\frac{\gamma}{\rho}}=\mathbb{E}_{\Inftagg^{\kappa}}\left\{\left(-\Inftagg^{\kappa}\right)^t\exp\left(-m\frac{\gamma}{\rho}\Inftagg^{\kappa}\right)\right\}.
\end{align}

Substituting~\eqref{eq:outage_de3} into~\eqref{eq:outage_de2}, we have
\begin{align}
P_{\out}^{\kappa}(\gamma)=1-\sum\limits_{t=0}\limits^{m-1}\frac{(-s)^{t}}{t!}\left.\frac{d^t}{d s^t}\mathcal{M}_{\Inftagg^{\kappa}}(s)\right|_{s=m\frac{\gamma}{\rho}}.
\end{align}
\end{proof}

\section{Derivation of Proposition~\ref{prop_schemeBS}: MGF of the interference at BS }\label{app_scheme_BS}
\begin{proof}
For the considered mode selection scheme, the p-DUE is in D2D mode if and only if \redcom{$\rho_{\!D}r_d^{\alpha_{\!D}}r_c^{-\alpha_{\!C}}<\xi$} (equivalently, $r_c>r_d^{\!\frac{\alpha_{\!D}}{\alpha_{\!C}}}\!\left(\!\frac{\rho_{\!D}}{\xi}\!\right)^{\!\frac{1}{\alpha_{\!C}}}\triangleq r_d'$). Defining $\tilde{\rd}\triangleq\textup{min}\left(\rd,\RM^{\frac{\alpha_{\!C}}{\alpha_{\!D}}}\left(\frac{\xi}{\rho_{\!D}}\right)^{\frac{1}{\alpha_{\!D}}}\right)$, we can then express $\InfB$ as
\ifCLASSOPTIONonecolumn
\begin{align}\label{eq:exp_I}
\InfB=\begin{cases}
g\rho_{\!D} r_{d}^{\alpha_{\!D}}r_{c}^{-\alpha_{\!C}}, &(r_d'\leq r_c<\RM,0\leq r_d<\tilde{\rd}) ;\\
0, &(0\leq r_c<r_d',0\leq r_d<\tilde{\rd});\\
0, &(0\leq r_c<\RM,\tilde{\rd}\leq r_d<\rd);
\end{cases}
\end{align}
\else
\begin{align}\label{eq:exp_I}
\InfB=\begin{cases}
g\rho_{\!D} r_{d}^{\alpha_{\!D}}r_{c}^{-\alpha_{\!C}}, &(r_d'\leq r_c<\RM,0\leq r_d<\tilde{\rd}) ;\\
0, &(0\leq r_c<r_d',0\leq r_d<\tilde{\rd});\\
0, &(0\leq r_c<\RM,\tilde{\rd}\leq r_d<\rd);
\end{cases}
\end{align}
\fi

\ifCLASSOPTIONonecolumn
Using the definition of MGF, we have
  \begin{align}\label{eq:mgf_derive1}
&\mathcal{M}_{\InfB}(s) =\!\int_{0}^{\tilde{\rd}}\!\!\!\int_{r_d'}^{\RM}\!\!\!\int_{0}^{\infty}\!\!\!\!\exp\left(-sg\rho_{\!D} r_{d}^{\alpha_{\!D}}r_{c}^{-\alpha_{\!C}}\right)\frg\frc \frd \,\textup{d}g\,\textup{d}r_c \,\textup{d}r_d \nonumber\\
&+\int_{0}^{\tilde{\rd}}\!\!\!\int_{0}^{r_d'}\!\!\!\int_{0}^{\infty}\!\!\!\! \frg\frc\frd \,\textup{d}g\,\textup{d}r_c \,\textup{d}r_d
+\int_{\tilde{\rd}}^{\rd}\!\!\!\int_{0}^{\RM}\!\!\!\!\int_{0}^{\infty}\!\!\!\! \frg\frc\frd \,\textup{d}g\,\textup{d}r_c \,\textup{d}r_d\nonumber\\
&=1-\int_0^{\tilde{\rd}}\!\!\left(\!\,_2F_1\left[1,\frac{2}{\alpha_{\!C}},1+\frac{2}{\alpha_{\!C}},-\frac{\RM^{\alpha_{\!C}}}{s\rho_{\!D}r_d^{\alpha_{\!D}}}\!\right]\!-\!\frac{r_d^{\!2\frac{\alpha_{\!D}}{\alpha_{\!C}}} \,_2F_1\!\left[\!1,\frac{2}{\alpha_{\!C}},1+\frac{2}{\alpha_{\!C}},-\frac{1}{s\xi}\!\right]}{\RM^2(\xi/\rho_{\!D})^{\frac{2}{\alpha_{\!C}}}}\!\right)\!\frd \,\textup{d}r_d\nonumber\\
 &=1\!+\!\frac{\,_2F_1\left[1,\frac{2}{\alpha_{\!C}};1+\frac{2}{\alpha_{\!C}};\frac{-1}{s\xi}\right]}{\rd^2\RM^2\tilde{\rd}^{\!\!-2-\frac{2\alpha_{\!D}}{\alpha_{\!C}}}(\xi/\rho_{\!D})^{\frac{2}{\alpha_{\!C}}}}\frac{\alpha_{\!C}}{\alpha_{\!D}+\alpha_{\!C}}\nonumber\\
 &\quad-\begin{cases}
\left.\left[\frac{\!\alpha_{\!C} \,_2F_1\left[1,\frac{2}{\alpha_{\!C}};1+\frac{2}{\alpha_{\!C}};\frac{-\RM^{\alpha_{\!C}}}{s\rho_{\!D} x^{\alpha_{\!D}}}\right]
\!+\!\alpha_{\!D} \,_2F_1\!\left[1,\frac{-2}{\alpha_{\!D}};1-\frac{2}{\alpha_{\!D}};\frac{-\RM^{\alpha_{\!C}}}{s\rho_{\!D} x^{\alpha_{\!D}}}\right]}{x^{-2}\rd^2(\alpha_{\!C}+\alpha_{\!D})}\right]\right|_0^{\tilde{\rd}}
,  &{\alpha_{\!D}\neq2}; \\
\frac{2\tilde{\rd}^2\textup{MeijerG}\left[\left\{\left\{0,\frac{\alpha_{\!C}-2}{\alpha_{\!C}}\right\},\{2\}\right\}, \left\{\left\{0,1\right\},\left\{\frac{-2}{\alpha_{\!C}}\right\}\right\},\frac{\RM^{\alpha_{\!C}}}{s\rho_{\!D}\tilde{\rd}^2}\right]}{\rd^2\alpha_{\!C}}
,  &{\alpha_{\!D}=2};
\end{cases}
  \end{align}
  \else
  Using the definition of MGF, we have~\eqref{eq:mgf_derive1} as shown at the top of next page,
    \begin{figure*}[!t]
\normalsize
  \begin{align}\label{eq:mgf_derive1}
&\mathcal{M}_{\InfB}(s) =\!\int_{0}^{\tilde{\rd}}\!\!\!\int_{r_d'}^{\RM}\!\!\!\int_{0}^{\infty}\!\!\!\!\exp\left(-sg\rho_{\!D} r_{d}^{\alpha_{\!D}}r_{c}^{-\alpha_{\!C}}\right)\frg\frc \frd \,\textup{d}g\,\textup{d}r_c \,\textup{d}r_d \nonumber\\
&+\int_{0}^{\tilde{\rd}}\!\!\!\int_{0}^{r_d'}\!\!\!\int_{0}^{\infty}\!\!\!\! \frg\frc\frd \,\textup{d}g\,\textup{d}r_c \,\textup{d}r_d
+\int_{\tilde{\rd}}^{\rd}\!\!\!\int_{0}^{\RM}\!\!\!\!\int_{0}^{\infty}\!\!\!\! \frg\frc\frd \,\textup{d}g\,\textup{d}r_c \,\textup{d}r_d\nonumber\\
&=1-\int_0^{\tilde{\rd}}\!\!\left(\!\,_2F_1\left[1,\frac{2}{\alpha_{\!C}},1+\frac{2}{\alpha_{\!C}},-\frac{\RM^{\alpha_{\!C}}}{s\rho_{\!D}r_d^{\alpha_{\!D}}}\!\right]\!-\!\frac{r_d^{\!2\frac{\alpha_{\!D}}{\alpha_{\!C}}} \,_2F_1\!\left[\!1,\frac{2}{\alpha_{\!C}},1+\frac{2}{\alpha_{\!C}},-\frac{1}{s\xi}\!\right]}{\RM^2(\xi/\rho_{\!D})^{\frac{2}{\alpha_{\!C}}}}\!\right)\!\frd \,\textup{d}r_d\nonumber\\
 &=1\!+\!\frac{\,_2F_1\left[1,\frac{2}{\alpha_{\!C}};1+\frac{2}{\alpha_{\!C}};\frac{-1}{s\xi}\right]}{\rd^2\RM^2\tilde{\rd}^{\!\!-2-\frac{2\alpha_{\!D}}{\alpha_{\!C}}}(\xi/\rho_{\!D})^{\frac{2}{\alpha_{\!C}}}}\frac{\alpha_{\!C}}{\alpha_{\!D}+\alpha_{\!C}}-\begin{cases}
\left.\left[\frac{\!\alpha_{\!C} \,_2F_1\left[1,\frac{2}{\alpha_{\!C}};1+\frac{2}{\alpha_{\!C}};\frac{-\RM^{\alpha_{\!C}}}{s\rho_{\!D} x^{\alpha_{\!D}}}\right]
\!+\!\alpha_{\!D} \,_2F_1\!\left[1,\frac{-2}{\alpha_{\!D}};1-\frac{2}{\alpha_{\!D}};\frac{-\RM^{\alpha_{\!C}}}{s\rho_{\!D} x^{\alpha_{\!D}}}\right]}{x^{-2}\rd^2(\alpha_{\!C}+\alpha_{\!D})}\right]\right|_0^{\tilde{\rd}}
,  &{\alpha_{\!D}\neq2}; \\
\frac{2\tilde{\rd}^2\textup{MeijerG}\left[\left\{\left\{0,\frac{\alpha_{\!C}-2}{\alpha_{\!C}}\right\},\{2\}\right\}, \left\{\left\{0,1\right\},\left\{\frac{-2}{\alpha_{\!C}}\right\}\right\},\frac{\RM^{\alpha_{\!C}}}{s\rho_{\!D}\tilde{\rd}^2}\right]}{\rd^2\alpha_{\!C}}
,  &{\alpha_{\!D}=2};
\end{cases}
  \end{align}
\hrulefill
\vspace*{4pt}
\vspace{-0.05 in}
\end{figure*}
\fi
  \noindent where the final result is obtained using~\cite{gradshteyn2007} and Mathematica software.
\end{proof}

\section{Derivation of Proposition~\ref{prop_scheme2}: MGF of the interference from p-DUE }\label{app_scheme2}
\begin{proof}
For a DRx located at distance $\dis$ away from the BS, the interference from an i.u.d. p-DUE, $\InfD$, is similar to~\eqref{eq:exp_I} except that $g\rho_{\!D} r_{d}^{\alpha_{\!D}}r_{c}^{-\alpha_{\!C}}$ is replaced by $g\rho_{\!D} r_d^{\alpha_{\!D}}\left(r_c^2+\dis^2-2r_c\dis\cos\theta\right)^{-\frac{\alpha_{\!D}}{2}}$, where $\theta$ is the angle formed between the $y'$-BS line and p-DUE-BS line, which is uniformly distributed between $0$ and $2\pi$ (see Fig.~\ref{fig:system}). Using the definition of MGF and simplifying, we have
\ifCLASSOPTIONonecolumn
\redcom{\begin{align}\label{eq:total_MFG1}
&\mathcal{M}_{\InfD}(s,\dis)=\int_{0}^{\tilde{\rd}}\!\!\!\!\int_{r_d'}^{\RM}\!\int_0^{2\pi}\!\!\!\int_{0}^{\infty}\!\!\!\!\exp\!\left(\!-sg\rho_{\!D} r_d^{\alpha_{\!D}}\!\left(\!r_c^2\!+\!\dis^2\!-\!2r_c\dis\cos\theta\right)^{\!-\frac{\alpha_{\!D}}{2}}\!\right)\frg\frac{1}{2\pi}\frc \frd \,\textup{d}g\,\textup{d}\theta\,\textup{d}r_c \,\textup{d}r_d \nonumber\\
&+\!\int_{0}^{\tilde{\rd}}\!\!\!\int_{0}^{r_d'}\!\int_0^{2\pi}\!\!\!\int_{0}^{\infty}\!\!\!\! \frg\frac{1}{2\pi}\frc\frd \,\textup{d}g\,\textup{d}\theta\,\textup{d}r_c\textup{d}r_d
\!+\!\!\int_{\tilde{\rd}}^{\rd}\!\!\!\int_{0}^{\RM}\!\int_0^{2\pi}\!\!\!\!\int_{0}^{\infty}\!\!\!\! \frg\frac{1}{2\pi}\frc\frd \,\textup{d}g\,\textup{d}\theta\,\textup{d}r_c \textup{d}r_d\nonumber\\
&=1-\!\!\int_{0}^{\tilde{\rd}}\!\!\!\int_{r_d'}^{\RM}\!\!\!\int_0^{\pi}\frac{s\rho_{\!D}r_d^{\alpha_{\!D}}}{s\rho_{\!D}r_d^{\alpha_{\!D}}\!+\!\left(r_c^2\!+\!\dis^2\!-\!2r_c\dis\cos\theta\right)^{\frac{\alpha_{\!D}}{2}}}\frac{1}{\pi}\frc\frd \,\textup{d}\theta \,\textup{d}r_c\,\textup{d}r_d,
\end{align}}
\else
\redcom{\begin{small}
\begin{align}\label{eq:total_MFG1}
&\mathcal{M}_{\InfD}\!(s,\dis)\!=\!\int_{0}^{\tilde{\rd}}\!\!\!\!\!\int_{r_d'}^{\RM}\!\!\int_0^{2\pi}\!\!\!\int_{0}^{\infty}\!\!\!\!\!\exp\!\left(\!\frac{-sg\rho_{\!D} r_d^{\alpha_{\!D}}}{\left(\!r_c^2\!+\!\dis^2\!-\!2r_c\dis\cos\theta\right)^{\!\frac{\alpha_{\!D}}{2}}}\!\right)\nonumber\\
&\quad\quad\times\frg\frac{1}{2\pi}\frc \frd \,\textup{d}g\,\textup{d}\theta\,\textup{d}r_c \,\textup{d}r_d \nonumber\\
&+\!\int_{0}^{\tilde{\rd}}\!\!\!\int_{0}^{r_d'}\!\int_0^{2\pi}\!\!\!\int_{0}^{\infty}\!\!\!\! \frg\frac{1}{2\pi}\frc\frd \,\textup{d}g\,\textup{d}\theta\,\textup{d}r_c\textup{d}r_d\nonumber\\
&\!+\!\!\int_{\tilde{\rd}}^{\rd}\!\!\!\int_{0}^{\RM}\!\int_0^{2\pi}\!\!\!\!\int_{0}^{\infty}\!\!\!\! \frg\frac{1}{2\pi}\frc\frd \,\textup{d}g\,\textup{d}\theta\,\textup{d}r_c \textup{d}r_d\nonumber\\
&\!=\!1-\!\!\int_{0}^{\tilde{\rd}}\!\!\!\int_{r_d'}^{\RM}\!\!\!\int_0^{\pi}\frac{s\rho_{\!D}r_d^{\alpha_{\!D}}}{s\rho_{\!D}r_d^{\alpha_{\!D}}\!+\!\left(r_c^2\!+\!\dis^2\!-\!2r_c\dis\cos\theta\right)^{\frac{\alpha_{\!D}}{2}}}\nonumber\\
&\quad\quad\times\frac{1}{\pi}\frc\frd \,\textup{d}\theta \,\textup{d}r_c\,\textup{d}r_d,
\end{align}
\end{small}}
\fi
\noindent where $r_d'\triangleq r_d^{\!\frac{\alpha_{\!D}}{\alpha_{\!C}}}\!\left(\!\frac{\rho_{\!D}}{\xi}\!\right)^{\!\frac{1}{\alpha_{\!C}}}$.

Due to the complicated expression of $\InfD$, the closed-form results (or semi-closed form) exist only for the cases of $\alpha_{\!D}=2$ or $\alpha_{\!D}=4$.

$\bullet$ {\underline{Case of $\alpha_{\!D}=2$:}}
\ifCLASSOPTIONonecolumn
Substituting $\alpha_{\!D}=2$ into~\eqref{eq:total_MFG1}, we get
\begin{subequations}\label{eq:mgf_derive2}
\begin{align}
&\mathcal{M}_{\InfD}(s,\dis)=1-\!\!\int_{0}^{\tilde{\rd}}\!\!\!\int_{r_d'}^{\RM}\!\!\!\int_{0}^{\pi}\frac{s\rho_{\!D}r_d^2}{s\rho_{\!D}r_d^2\!+\!r_c^2\!+\!\dis^2\!-\!2r_c\dis\cos\theta}\frac{1}{\pi}\frc\frd\,\textup{d}\theta \,\textup{d}r_c\,\textup{d}r_d    \nonumber\\
&=1-\!\!\int_{0}^{\tilde{\rd}}\!\!\!\int_{r_d'}^{\RM}\!\frac{s\rho_{\!D}r_d^2}{\sqrt{(s\rho_{\!D}r_d^2+r_c^2+\dis^2)^2-4r_c^2\dis^2}}\frac{2r_c}{\RM^2}\frd \,\textup{d}r_c \,\textup{d}r_d  \quad\quad\quad \nonumber\\
&=\!1\!-\!\!\int_{0}^{\tilde{\rd}}\!\!\frac{s\rho_{\!D}r_d^2}{\RM^2}\ln\left(\frac{\beta_1\left(r_d^2,s\rho_{\!D},\RM^2-\dis^2,4\dis^2s\rho_{\!D}\right)}
{\!\sqrt{\!\left(\!\left(\frac{r_d^2\rho_{\!D}}{\xi}\right)^{\!\frac{2}{\alpha_{\!C}}}\!+\!s\rho_{\!D}r_d^2\!-\!\dis^2\!\right)^{\!2}\!+\!4\dis^2s\rho_{\!D}r_d^2}\!+\!\left(\frac{r_d^2\rho_{\!D}}{\xi}\right)^{\!\frac{2}{\alpha_{\!C}}}\!+\!s\rho_{\!D}r_d^2\!-\!\dis^2}\right)\frac{2r_d}{\rd^2}\,\textup{d}r_d \label{eq:mgf_lterm_other2}\\
&=\!1\!-\!\frac{s\rho_{\!D}\left.\left[\Psi_1\left(x^2,s\rho_{\!D},\RM^2-\dis^2,4\dis^2s\rho_{\!D}\right)-\Psi_1\left(x^2,s\rho_{\!D}+\frac{\rho_{\!D}}{\xi},-\dis^2,4\dis^2s\rho_{\!D}\right)\right]\right|_{0}^{\tilde{\rd}}}{\rd^2\RM^2}, \quad(\alpha_{\!C}=2),
\end{align}
\end{subequations}
\else
Substituting $\alpha_{\!D}=2$ into~\eqref{eq:total_MFG1}, we get~\eqref{eq:mgf_derive2} as shown at the top of next page,
  \begin{figure*}[!t]
\normalsize
\begin{subequations}\label{eq:mgf_derive2}
\begin{align}
&\mathcal{M}_{\InfD}(s,\dis)=1-\!\!\int_{0}^{\tilde{\rd}}\!\!\!\int_{r_d'}^{\RM}\!\!\!\int_{0}^{\pi}\frac{s\rho_{\!D}r_d^2}{s\rho_{\!D}r_d^2\!+\!r_c^2\!+\!\dis^2\!-\!2r_c\dis\cos\theta}\frac{1}{\pi}\frc\frd\,\textup{d}\theta \,\textup{d}r_c\,\textup{d}r_d    \nonumber\\
&=1-\!\!\int_{0}^{\tilde{\rd}}\!\!\!\int_{r_d'}^{\RM}\!\frac{s\rho_{\!D}r_d^2}{\sqrt{(s\rho_{\!D}r_d^2+r_c^2+\dis^2)^2-4r_c^2\dis^2}}\frac{2r_c}{\RM^2}\frd \,\textup{d}r_c \,\textup{d}r_d  \quad\quad\quad \nonumber\\
&=\!1\!-\!\!\int_{0}^{\tilde{\rd}}\!\!\frac{s\rho_{\!D}r_d^2}{\RM^2}\ln\left(\frac{\beta_1\left(r_d^2,s\rho_{\!D},\RM^2-\dis^2,4\dis^2s\rho_{\!D}\right)}
{\!\sqrt{\!\left(\!\left(\frac{r_d^2\rho_{\!D}}{\xi}\right)^{\!\frac{2}{\alpha_{\!C}}}\!+\!s\rho_{\!D}r_d^2\!-\!\dis^2\!\right)^{\!2}\!+\!4\dis^2s\rho_{\!D}r_d^2}\!+\!\left(\frac{r_d^2\rho_{\!D}}{\xi}\right)^{\!\frac{2}{\alpha_{\!C}}}\!+\!s\rho_{\!D}r_d^2\!-\!\dis^2}\right)\frac{2r_d}{\rd^2}\,\textup{d}r_d \label{eq:mgf_lterm_other2}\\
&=\!1\!-\!\frac{s\rho_{\!D}\left.\left[\Psi_1\left(x^2,s\rho_{\!D},\RM^2-\dis^2,4\dis^2s\rho_{\!D}\right)-\Psi_1\left(x^2,s\rho_{\!D}+\frac{\rho_{\!D}}{\xi},-\dis^2,4\dis^2s\rho_{\!D}\right)\right]\right|_{0}^{\tilde{\rd}}}{\rd^2\RM^2}, \quad(\alpha_{\!C}=2).
\end{align}
\end{subequations}
\hrulefill
\vspace*{4pt}
\vspace{-0.15 in}
\end{figure*}
\fi
\noindent where the second and third steps come from (2.553) and (2.261) in~\cite{gradshteyn2007}, respectively, and last step is obtained using Mathematica. $\beta_1(x,a,b,c)=ax+b+\sqrt{(ax+b)^2+cx}$, $\int_{x}x\beta_1(x,a,b,c)\textup{d}x=\Psi_1(x,a,b,c)$ and
\ifCLASSOPTIONonecolumn
 \begin{align}\label{eq:psi1}
\Psi_1(x,a,b,c)&=\frac{-x^2}{8}+\frac{(10ab+3c-2a^2x)\sqrt{(ax+b)^2+cx}}{16}+\frac{x^2}{2}\ln\left(\beta_1(x,a,b,c)\right)\nonumber\\
&-\frac{(16a^2b^2+16abc+3c^2)\ln\left(c+2a^2x+2a\left(b+\sqrt{(ax+b)^2+cx}\right)\right)}{32a^4}.
 \end{align}
 \else
  \begin{align}\label{eq:psi1}
&\Psi_1(x,a,b,c)=\frac{-x^2}{8}+\frac{(10ab+3c-2a^2x)\sqrt{(ax+b)^2+cx}}{16}\nonumber\\
&\!+\!\frac{x^2}{2}\!\ln\!\left(\beta_1(x,a,b,c)\right)\!-\!\frac{\ln\!\left(\!c\!+\!2a^2x\!+\!2a\!\left(\!b\!+\!\sqrt{(ax\!+\!b)^2\!+\!cx}\right)\!\right)}{32a^4(16a^2b^2+16abc+3c^2)^{-1}}.
 \end{align}
 \fi

$\bullet$ {\underline{Case of $\alpha_{\!D}=4$:}}
\ifCLASSOPTIONonecolumn
Similar to $\alpha_{\!D}=2$ case, via substituting $\alpha_{\!D}=4$ into~\eqref{eq:total_MFG1} and using (2.553) and (2.261) in~\cite{gradshteyn2007}, we have
\begin{small}
\begin{subequations}
\begin{align}
&\mathcal{M}_{\InfD}(s,\dis)\!=\!1\!-\!\!\int_{0}^{\tilde{\rd}}\!\!\!\!\!\int_{r_d'}^{\RM}\!\!\!\!\!\int_{0}^{\pi}\!\!\frac{s\rho_{\!D}r_d^4}{2\im\sqrt{s\rho_{\!D}r_d^4}}\!\left(\!
 \frac{1}{r_c^2\!+\!\dis^2\!-\!2r_c\dis\cos\theta\!-\!\im\sqrt{s\rho_{\!D}r_d^4}}-\frac{1}{r_c^2\!+\!\dis^2\!-\!2r_c\dis\cos\theta\!+\!\im\sqrt{s\rho_{\!D}r_d^4}}
 \!\right)\!\!\frac{2r_c\,\textup{d}\theta \,\textup{d}r_c}{\pi\RM^2}\frd\,\textup{d}r_d\nonumber\\
&=\!1\!-\!\!\int_{0}^{\tilde{\rd}}\!\!\!\frac{\sqrt{s\rho_{\!D}r_d^4}}{2\im \RM^2}\!\int_{r_d'}^{\RM}\!\!\!\left(\frac{2r_c}{\sqrt{(r_c^2\!+\!\dis^2\!-\!\im\sqrt{s\rho_{\!D}r_d^4})^2\!-\!4r_c^2\dis^2}}-\frac{2r_c}{\sqrt{(r_c^2\!+\!\dis^2\!+\!\im\sqrt{s\rho_{\!D}r_d^4})^2\!-\!4r_c^2\dis^2}}    \right)\,\textup{d}r_c \frd\,\textup{d}r_d \nonumber\\
&=\!1\!-\!\!\int_{0}^{\!\tilde{\rd}}\!\!\frac{\sqrt{s\rho_{\!D}r_d^4}}{\RM^2}\textup{Im}\!\!\left\{\!\!\ln
\frac{\beta_1\left(r_d^2,-\im\sqrt{s\rho_{\!D}},\RM^2\!-\!\dis^2,-4\im\sqrt{s\rho_{\!D}}\dis^2\right)}
{\!\sqrt{\!\left(\!\left(\!\frac{r_d^4\rho_{\!D}}{\xi}\!\right)^{\!\frac{2}{\alpha_{\!C}}}\!\!-\!\im\sqrt{s\rho_{\!D}}r_d^2\!-\!\dis^2\!\right)^{\!2}\!-\!4\im\sqrt{s\rho_{\!D}}\dis^2r_d^2}\!+\left(\!\frac{r_d^4\rho_{\!D}}{\xi}\!\right)^{\!\frac{2}{\alpha_{\!C}}}\!\!-\!\im\sqrt{s\rho_{\!D}}r_d^2\!-\!\dis^2}\!\right\}\!
\frac{2r_d}{\rd^2}\,\textup{d}r_d \label{eq:mgf_lterm_other4}\\
&=\!1\!-\!\frac{\textup{Im}\left\{\!\left.\left[\Psi_1\!\left(x^2,-\im\sqrt{s\rho_{\!D}},\RM^2\!-\!\dis^2,-4\im\sqrt{s\rho_{\!D}}\dis^2\right)
\!-\!\Psi_1\!\left(x^2,\sqrt{\frac{\rho_{\!D}}{\xi}}\!-\!\im\sqrt{s\rho_{\!D}},-\dis^2,-4\im\sqrt{s\rho_{\!D}}\dis^2\right)\right]\right|_{0}^{\tilde{\rd}}\!\right\}}{(\sqrt{s\rho_{\!D}})^{-1}\rd^2\RM^2},\quad(\alpha_{\!C}=4),
\end{align}
\end{subequations}
\end{small}
\else
  Similar to $\alpha_{\!D}=2$ case, via substituting $\alpha_{\!D}=4$ into~\eqref{eq:total_MFG1} and using (2.553) and (2.261) in~\cite{gradshteyn2007}, we have~\eqref{eq:mgf_derive3} as shown at the top of next page,
  \begin{figure*}[!t]
\normalsize
\begin{small}
\begin{subequations}\label{eq:mgf_derive3}
\begin{align}
&\mathcal{M}_{\InfD}(s,\dis)\!=\!1\!-\!\!\int_{0}^{\tilde{\rd}}\!\!\!\!\!\int_{r_d'}^{\RM}\!\!\!\!\!\int_{0}^{\pi}\!\!\frac{s\rho_{\!D}r_d^4}{2\im\sqrt{s\rho_{\!D}r_d^4}}\!\left(\!
 \frac{1}{r_c^2\!+\!\dis^2\!-\!2r_c\dis\cos\theta\!-\!\im\sqrt{s\rho_{\!D}r_d^4}}-\frac{1}{r_c^2\!+\!\dis^2\!-\!2r_c\dis\cos\theta\!+\!\im\sqrt{s\rho_{\!D}r_d^4}}
 \!\right)\!\!\frac{2r_c\,\textup{d}\theta \,\textup{d}r_c}{\pi\RM^2}\frd\,\textup{d}r_d\nonumber\\
&=\!1\!-\!\!\int_{0}^{\tilde{\rd}}\!\!\!\frac{\sqrt{s\rho_{\!D}r_d^4}}{2\im \RM^2}\!\int_{r_d'}^{\RM}\!\!\!\left(\frac{2r_c}{\sqrt{(r_c^2\!+\!\dis^2\!-\!\im\sqrt{s\rho_{\!D}r_d^4})^2\!-\!4r_c^2\dis^2}}-\frac{2r_c}{\sqrt{(r_c^2\!+\!\dis^2\!+\!\im\sqrt{s\rho_{\!D}r_d^4})^2\!-\!4r_c^2\dis^2}}    \right)\,\textup{d}r_c \frd\,\textup{d}r_d \nonumber\\
&=\!1\!-\!\!\int_{0}^{\!\tilde{\rd}}\!\!\frac{\sqrt{s\rho_{\!D}r_d^4}}{\RM^2}\textup{Im}\!\!\left\{\!\!\ln
\frac{\beta_1\left(r_d^2,-\im\sqrt{s\rho_{\!D}},\RM^2\!-\!\dis^2,-4\im\sqrt{s\rho_{\!D}}\dis^2\right)}
{\!\sqrt{\!\left(\!\left(\!\frac{r_d^4\rho_{\!D}}{\xi}\!\right)^{\!\frac{2}{\alpha_{\!C}}}\!\!-\!\im\sqrt{s\rho_{\!D}}r_d^2\!-\!\dis^2\!\right)^{\!2}\!-\!4\im\sqrt{s\rho_{\!D}}\dis^2r_d^2}\!+\left(\!\frac{r_d^4\rho_{\!D}}{\xi}\!\right)^{\!\frac{2}{\alpha_{\!C}}}\!\!-\!\im\sqrt{s\rho_{\!D}}r_d^2\!-\!\dis^2}\!\right\}\!
\frac{2r_d}{\rd^2}\,\textup{d}r_d \label{eq:mgf_lterm_other4}\\
&=\!1\!-\!\frac{\textup{Im}\left\{\!\left.\left[\Psi_1\!\left(x^2,-\im\sqrt{s\rho_{\!D}},\RM^2\!-\!\dis^2,-4\im\sqrt{s\rho_{\!D}}\dis^2\right)
\!-\!\Psi_1\!\left(x^2,\sqrt{\frac{\rho_{\!D}}{\xi}}\!-\!\im\sqrt{s\rho_{\!D}},-\dis^2,-4\im\sqrt{s\rho_{\!D}}\dis^2\right)\right]\right|_{0}^{\tilde{\rd}}\!\right\}}{(\sqrt{s\rho_{\!D}})^{-1}\rd^2\RM^2},\quad(\alpha_{\!C}=4).
\end{align}
\end{subequations}
\end{small}
\hrulefill
\vspace*{4pt}
\vspace{-0.05 in}
\end{figure*}
\fi
\noindent where the third step comes from the fact that the two integrated terms in the second step are conjugated such that $\frac{a-a*}{2\im}=\textup{Im}\{a\}$.

Thus, we arrive at the result in Proposition~\ref{prop_scheme2}.
\end{proof}

\section{Derivation of Corollary~\ref{prop_CUE}: MGF of the interference from CUE }\label{app_coro1}
\begin{proof}
Since there is no constraint on the CUE, the i.u.d. CUE will always generate interference (e.g., $\Inftcue=g\rho_{\BS} r_z^{\alpha_{\!C}}\left(r_z^2+\dis^2-2r_z\dis\cos\theta\right)^{-\frac{\alpha_{\!D}}{2}}$) to this typical DRx. As before, we can only derive the analytical result for $\alpha_{\!D}=2$ or $4$.

$\bullet$ {\underline{Case of $\alpha_{\!D}=2$:}}
According to the definition of MGF and the expression of $\Inftcue$, we have
\ifCLASSOPTIONonecolumn
 \begin{subequations}
\begin{align}
\mathcal{M}_{\Inftcue}(s,\dis)&=1-\int_{0}^{\RM}\!\!\!\int_0^{\pi}\frac{s\rho_{\BS}r_z^{\alpha_{\!C}}}{s\rho_{\BS}r_z^{\alpha_{\!C}}\!+\!\left(r_z^2\!+\!\dis^2\!-\!2r_z\dis\cos\theta\right)^{\frac{\alpha_{\!D}}{2}}}\frac{1}{\pi}\frac{2r_z}{\RM^2} \,\textup{d}\theta \,\textup{d}r_z\label{eq:mgf_cue_step1}\\
&=1-\frac{s\rho_{\BS}}{\RM^2}\int_{0}^{\RM}\frac{2r_z^{\alpha_{\!C}+1}}{\sqrt{(s\rho_{\BS}r_z^{\alpha_{\!C}}+r_z^2+\dis^2)^2-4r_z^2\dis^2}}\,\textup{d}r_z \label{eq:mgf_cue_2} \\
&=1-\frac{s\rho_{\BS}\left.\left[\beta_2\left(x^2,(s\rho_{\BS}+1)^2,\dis^2(s\rho_{\BS}-1),4\dis^4s\rho_{\BS}\right)\right]\right|_{0}^{\RM}}{\RM^2(s\rho_{\BS}+1)^3}, \quad{(\alpha_{\!C}=2),}\label{eq:mgf_cue_step3}
\end{align}
\end{subequations}
\else
\begin{small}
 \begin{subequations}
\begin{align}
&\mathcal{M}_{\Inftcue}(s,\dis)\!=1\!-\!\!\int_{0}^{\RM}\!\!\!\int_0^{\pi}\!\!\!\frac{s\rho_{\BS}r_z^{\alpha_{\!C}}2r_z/(\pi\RM^2)}{s\rho_{\BS}r_z^{\alpha_{\!C}}\!+\!\left(r_z^2\!+\!\dis^2\!-\!2r_z\dis\cos\theta\right)^{\frac{\alpha_{\!D}}{2}}} \,\textup{d}\theta \,\textup{d}r_z\label{eq:mgf_cue_step1}\\
&=1-\frac{s\rho_{\BS}}{\RM^2}\int_{0}^{\RM}\frac{2r_z^{\alpha_{\!C}+1}}{\sqrt{(s\rho_{\BS}r_z^{\alpha_{\!C}}+r_z^2+\dis^2)^2-4r_z^2\dis^2}}\,\textup{d}r_z \label{eq:mgf_cue_2} \\
&=1-\frac{s\rho_{\BS}\left.\left[\beta_2\left(x^2,(s\rho_{\BS}+1)^2,\dis^2(s\rho_{\BS}-1),4\dis^4s\rho_{\BS}\right)\right]\right|_{0}^{\RM}}{\RM^2(s\rho_{\BS}+1)^3}, \nonumber\\&  \quad \quad \quad \quad \quad \quad \quad \quad \quad \quad \quad \quad \quad \quad \quad \quad \quad \quad \quad \quad \quad\quad{(\alpha_{\!C}=2),}\label{eq:mgf_cue_step3}
\end{align}
\end{subequations}
\end{small}
\fi
where $\beta_2(x,a,b,c)=\sqrt{(ax+b)^2+c}-b\ln\left(ax+b+\sqrt{(ax+b)^2+c}\right)$.

$\bullet$ {\underline{Case of $\alpha_{\!D}=4$:}}
Similarly, substituting $\alpha_{\!D}=4$ into~\eqref{eq:mgf_cue_step1}, we get
\ifCLASSOPTIONonecolumn
 \begin{subequations}
\begin{align}
\mathcal{M}_{\Inftcue}(s,\dis)
&=1-\int_{0}^{\RM}\textup{Im}\!\left\{\frac{r_z^{\alpha_{\!C}/2}}{\sqrt{\left(r_z^2+\dis^2-\im\sqrt{s\rho_{\BS}r_z^{\alpha_{\!C}}}\right)^2-4r_z^2\dis^2}}\right\}\frac{2r_z\sqrt{s\rho_{\BS}}}{\RM^2} \,\textup{d}r_z \label{eq:mgf_cue_4}  \\
&=1-\textup{Im}\left\{\frac{\sqrt{s\rho_{\BS}}\left[\left.\beta_2\left(x^2,1-\im\sqrt{s\rho_{\BS}},-\dis^2\frac{1+\im\sqrt{s\rho_{\BS}}}{1-\im\sqrt{s\rho_{\BS}}},\frac{-4\im\sqrt{s\rho_{\BS}}\dis^4}{(1-\im\sqrt{s\rho_{\BS}})^2}\right)\right]\right|_{0}^{\RM}}{\RM^2(1-\im\sqrt{s\rho_{\BS}})^2}\right\}, \quad{(\alpha_{\!C}=4)}\label{eq:mgf_cue_step5}.
\end{align}
\end{subequations}
\else
\begin{small}
 \begin{subequations}
\begin{align}
&\mathcal{M}_{\Inftcue}(s,\dis)
\!=1\!-\!\!\int_{0}^{\RM}\!\!\textup{Im}\!\left\{\frac{r_z^{\alpha_{\!C}/2}}{\sqrt{\left(r_z^2\!+\!\dis^2\!-\!\im\sqrt{s\rho_{\BS}r_z^{\alpha_{\!C}}}\right)^2\!\!-\!4r_z^2\dis^2}}\right\}\nonumber\\
&\quad\quad\quad\quad\quad\quad\quad\quad\quad\quad\quad\quad\quad\quad\quad\times\frac{2r_z\sqrt{s\rho_{\BS}}}{\RM^2} \,\textup{d}r_z \label{eq:mgf_cue_4}  \\
&\!=\!1\!-\!\textup{Im}\!\left\{\!\!\!\frac{\sqrt{s\rho_{\BS}}\left[\left.\!\beta_2\!\!\left(\!x^2,1\!-\!\im\sqrt{s\rho_{\BS}},\!-\dis^2\frac{1+\im\sqrt{s\rho_{\BS}}}{1-\im\sqrt{s\rho_{\BS}}},\frac{-4\im\sqrt{s\rho_{\BS}}\dis^4}{(1-\im\sqrt{s\rho_{\BS}})^2}\!\right)\!\right]\!\!\right|_{0}^{\RM}}{\RM^2(1-\im\sqrt{s\rho_{\BS}})^2}\!\!\right\},\nonumber\\
&\quad\quad\quad\quad\quad\quad\quad\quad\quad\quad\quad\quad\quad\quad\quad\quad \quad\quad\quad{(\alpha_{\!C}=4)}\label{eq:mgf_cue_step5}.
\end{align}
\end{subequations}
\end{small}
\fi

Note the step in~\eqref{eq:mgf_cue_step3} and step in~\eqref{eq:mgf_cue_step5} come from~\cite[(2.264)]{gradshteyn2007}.
\end{proof}
\section{Derivation of Proposition~\ref{prop:link}: Average Number of Successful D2D Transmissions }\label{app_link}
\begin{proof}
Using the definition in Section~\ref{subsec:spectraleff}, the average number of successful D2D transmissions can be mathematically written as
  \begin{align}\label{eq:spectral1}
   \bar{M} =& \mathbb{E}_{\totalphi}\left\{\sum\limits_{x_{i}\in\totalphi}\textbf{1}(x_j\!\in\!\phiDUE\!)\textbf{1}(\textup{SIR}^{\DRx}(x_j,y_j)\!>\!\gamma)\right\}.
  \end{align}

 As mentioned in Section~\ref{sec:MGF-DRx}, the location of underlay DRxs (i.e., whose corresponding p-DUE is in underlay D2D mode) follows the PPP. According to the reduced Campbell measure~\cite{Haenggi-2012}, we can rewrite~\eqref{eq:spectral1} as
 \begin{align}\label{eq:proofllink}
\bar{M}&=\int_{\area}\textbf{1}(x'\!\in\!\phiDUE)\textup{Pr}^{!}_{x'}\!\!\left(\textup{SIR}^{\DRx}(x',y')\!>\!\gamma\right) \densityD \,\textup{d}x'\nonumber\\
&=\int_{\area^{\DRx}}\textbf{1}(y'\!\in\!\phiD_u)\left(1-P_{\out}^{\DRx}(\gamma,y')\right) \densityR(y') \,\textup{d}y'\nonumber\\
&=\int_0^{\RM+\rd}\!\!\!\! p_{\DD}(\dis) \left(1-P_{\out}^{\DRx}(\gamma,\dis)\right)\densityR(\dis)2\pi \dis \,\textup{d}\dis,
\end{align}
\noindent where $\textup{Pr}^{!}_x(\cdot)$ is the reduced Palm distribution, $\area^{\DRx}$ is the network region for DRxs (e.g., a disk region with radius $\RM+\rd$). The second step in~\eqref{eq:proofllink} results from the Slivnyak's theorem and the fact that we interpret this reduced Campbell measure from the point view of DRx, while the last step in~\eqref{eq:proofllink} is based on the isotropic property of the underlay network region and the independent thinning property of PPP.
\end{proof}
\section{Derivation of Proposition~\ref{theorem_density}: Node density of DRxs}\label{app_Density}
\begin{proof}
\begin{figure*}
\begin{minipage}[t]{0.5\linewidth}
\centering
\begin{tikzpicture}[scale=0.4 ]
\tikzstyle{every node}=[font=\small]
\draw[thick]  (0,0) circle (6);
\draw[thick]  (0,0) circle (6);
\fill[red] (0,0.5) -- (-0.35,-0.4) -- (0.6,-0.4) -- (0,0.55);

\draw[thick]  (-6.2,1.45) circle( 2);

\draw[thick] (3.5,0) circle( 2);
\draw[thick,blue](2.1,1.5) -- (1.8,-1);
\draw[thick,blue] (2.7,1.8) -- (2.3,-1.55);
\draw[thick,blue](3.35,2) -- (2.85,-1.95);
\draw[thick,blue] (4,1.95) -- (3.5,-2.05);
\draw[thick,blue] (4.55,1.65) -- (4.1,-1.9);
\draw[thick,blue] (5.1,1.2) -- (4.75,-1.65);

\draw[thick,blue] (-4.25,1.95) -- (-4.65,0.2);
\draw[thick,blue](-5.15,-0.3) -- (-4.5,2.5);
\draw[thick,blue] (-5.75,-0.5) -- (-4.9,2.95);
\draw[thick,blue] (-5.4,2.6) -- (-6,0);

\draw[dashed,->] (0,0) -- (-6.2,1.45);
\draw(-0.65,-4) node[above]{$\RM$};
\draw[dashed] (0,0) -- (0,-5.9);
\draw(-2.25,0.7) node[above]{$\dis$};
\draw[dashed] (0,0) -- (3.5,0);
\draw(1,0.5) node[above]{$\dis$};
\draw [dashed,->](-6.2,1.45) -- (-7,3.2);
\draw(-7.05,1.65) node[above]{$\rd$};
\draw(-4,-2.65) node[above]{$\mathcal{B}_1=\phi(\dis,\RM,\rd)$};
\draw[->] (-7.05,-1.65) -- (-5,0.5);
\draw(2,2.8) node[above]{$\mathcal{B}_1=\pi \rd^2$};
\draw[->] (2.1,2.85) -- (2.65,1.1);
\end{tikzpicture}
               \vspace{-0.05 in}
\caption{Illustration of results in Proposition~\ref{theorem_density}.}\label{fig:density}
\end{minipage}%
\begin{minipage}[t]{0.5\linewidth}
\centering
\begin{tikzpicture}[scale=0.5 ]
\tikzstyle{every node}=[font=\small]
\draw[thick]  (-4,0) circle (3);
\fill[red] (0,0.5) -- (-0.35,-0.4) -- (0.6,-0.4) -- (0,0.55);

\draw [->](-8,0) -- (4,0);
\draw [->](-4,-4.95) -- (-4,4.5);
\draw(-5,4) node[above]{$x$};
\draw(7.5,0) node[above]{$y$};
\draw[thick] (2,0) circle (4.5);
\draw[thick,blue] (-2.3,1.35) -- (-2.3,-1.35);
\draw[thick,blue] (-2.05,2) -- (-2.05,-2);
\draw[thick,blue] (-1.8,2.05) -- (-1.8,-2.05);
\draw[thick,blue] (-1.55,1.85) -- (-1.55,-1.85);
\draw[thick,blue] (-1.3,1.35) -- (-1.3,-1.35);

\draw(-0.45,-2.45) node[above]{$(\dis,0)$};
\draw[->] (0,-1.5) -- (0,-0.5);
\fill  (2,0) circle (0.1 );
\draw(2.05,-4.1) node[above]{$\left(\frac{\xi^{\frac{2}{\alpha}}\dis}{\xi^{\frac{2}{\alpha}}-\rho_{\!D}^{\frac{2}{\alpha}}},0\right)$};
\draw[->] (2,-2.5) -- (2,-0.2);

\draw[dashed,->] (2,0) -- (4,4);
\draw(2.2,1.5) node[above]{$\frac{\xi^{\frac{1}{\alpha}}\rho_{\!D}^{\frac{1}{\alpha}}\dis}{\xi^{\frac{2}{\alpha}}-\rho_{\!D}^{\frac{2}{\alpha}}}$};

\fill  (6.5,0) circle (0.1 );
\draw(5.25,-2.6) node[above]{$\left(\frac{\xi^{\frac{1}{\alpha}}\dis}{\xi^{\frac{1}{\alpha}}-\rho_{\!D}^{\frac{1}{\alpha}}},0\right)$};
\draw[->] (5.5,-1) -- (6.4,-0.2);

\fill  (-2.5,0) circle (0.1 );
\draw(-2.2,-4.7) node[above]{$\left(\frac{\xi^{\frac{1}{\alpha}}\dis}{\xi^{\frac{1}{\alpha}}+\rho_{\!D}^{\frac{1}{\alpha}}},0\right)$};
\draw[->] (-2.5,-3) -- (-2.5,-0.2);

\draw[dashed,->] (-4,0) -- (-5.5,2.5);
\draw(-5.5,1) node[above]{$R_D$};
\draw(-2,4) node[above]{$\mathcal{B}_2$};
\draw[->] (-2,3.5) -- (-1.9,1.5);
\end{tikzpicture}
               \vspace{-0.05 in}
\caption{Illustration of results in Proposition~\ref{prop_active}.}
\label{fig:active2}
\end{minipage}
\vspace{-0.05 in}
\end{figure*}

Rather than considering that there is a DRx uniformly distributed around the p-DUE, we can consider that for each DRx, there is a p-DUE which is uniformly distributed inside the disk region formed around DRx. If the network region is infinite, the p-DUE's node density inside the region $\pi\rd^2$ is $\densityD$. As a result, the node density for DRx is $\densityD$.

However, since we consider a finite region (i.e., a disk region), the p-DUE's node density is no longer $\densityD$ at certain locations (e.g., the cell edge). Hence, the DRx's node density is not $\densityD$. Instead, the node density becomes $\densityD\frac{\mathcal{B}_1}{\pi\rd^2}$, which depends on the location of DRx, where $\mathcal{B}_1$ denotes the overlap region between the cell network region $\pi \RM^2$ and the disk region $\pi\rd^2$ centered at the DRx which is $\dis$ away from BS.

As illustrated in Fig.~\ref{fig:density}, when $\dis\in [0,\RM-\rd)$, the disk region formed around DRx is always inside the network region. That is to say, $\mathcal{B}_1$ is always $\pi\rd^2$. Thus, we have $\densityR(\dis)=\densityD$ within the considered range. However, for the case that $\dis\in[\RM-\rd,\RM+\rd)$, the $\mathcal{B}_1$ becomes $\psi(\dis,\RM,\rd)$, where $\psi(\dis,\RM,\rd)$ is the overlap region formed by two disk with radii $\RM$ and $\rd$ which is separated by distance $\dis$ and its formulation is presented in~\eqref{eq:psi} in Lemma 1. Then we have $\densityR(\dis)=\densityD\frac{\psi(\dis,\RM,\rd)}{\pi\rd^2}$. For the rest of range (i.e., $\dis\geq\RM+\rd$), $\densityR(\dis)=0$.
Hence, we arrive at the result in Proposition~\ref{theorem_density}.
\end{proof}

\section{Derivation of Proposition~\ref{prop_active}: the probability of being in D2D mode}\label{app_active2}
\begin{proof}
Assume that a DRx is located at distance $\dis$ away from the BS. Similar to the derivation of Proposition~\ref{theorem_density}, we consider that, for this DRx, there is a p-DUE uniformly surrounding it\footnote{In fact, at the cell edge, the possible location of p-DUE is no longer a disk region. In this analysis, we consider the case where the radius of the network region is large compared to $\rd$ such that, for those DRx in the range of $[\RM-\rd, \RM+\rd]$, $p_{\DD}(\dis)=1$. However, our result can be easily extended to the other possible scenarios. Due to the space limitation, we do not show those results here. }.

According to the considered mode selection scheme, this DRx is in underlay if its corresponding p-DUE satisfies $\rho_{\!D} r_{d}^{\alpha_{\!D}}<\xi r_{c}^{\alpha_{\!C}}$. Due to the analytical complexity, we can only find the exact result for the same path-loss case, while an approximate result can be derived for different path-loss values.
\subsection{Same path-loss exponent}
Let us consider the case $\alpha_{\!C}=\alpha_{\!D}\triangleq\alpha$. Note that the maximum range for $r_d$ is $\rd$, while the minimum range of $r_c$ is $\textup{max}\left(0,\dis-\rd\right)$. Assuming $\dis>\rd$, if $\rho_{\!D}\rd^{\alpha}$ is less than $\xi\left(\dis-\rd\right)^{\alpha}$ (equivalently, $\dis\!\geq\!\left(\!1\!+\!\left(\!\frac{\rho_{\!D}}{\xi}\!\right)^{\!\frac{1}{\alpha}}\!\right)\!\rd$), the probability that a p-DUE is in D2D mode is $1$. Because for any possible location of p-DUE in the disk region centered at DRx, the p-DUE's distance to the BS is always greater than $\rd$ (i.e., the p-DUE's maximum distance to its DRx). Consequently, for the case that $\dis\!\geq\!\left(\!1\!+\!\left(\!\frac{\rho_{\!D}}{\xi}\!\right)^{\!\frac{1}{\alpha}}\!\right)\!\rd$, $p_{\DD}(\dis)$ is always 1.

Under the scenario that $\dis\!<\!\left(\!1\!+\!\left(\!\frac{\rho_{\!D}}{\xi}\!\right)^{\!\frac{1}{\alpha}}\!\right)\!\rd$, the analysis is more complicated. Let us consider the case of $\xi>\rho_{\!D}$. The location of DRx is assumed to be at the origin and the BS is $\dis$ away from the DRx. For example, the coordinate of BS is $(\dis,0)$, as shown in Fig.~\ref{fig:active2}. Let $(x,y)$ denote the coordinate of p-DUE. This p-DUE is not in D2D mode if the following requirement is met, i.e., $\rho_{\!D}\left(x^2+y^2\right)^{\frac{\alpha}{2}}>\xi\left((\dis-x)^2+y^2\right)^{\frac{\alpha}{2}}$. Note that $r_d=\sqrt{x^2+y^2}$ and $r_c=\sqrt{(\dis-x)^2+y^2}$. After rearranging this inequality, we have
\ifCLASSOPTIONonecolumn
\begin{align}
\left(x-\frac{\xi^{\frac{2}{\alpha}}\dis}{\xi^{\frac{2}{\alpha}}-\rho_{\!D}^{\frac{2}{\alpha}}}\right)^2+y^2<\left(\frac{\xi^{\frac{1}{\alpha}}\rho_{\!D}^{\frac{1}{\alpha}}\dis}{\xi^{\frac{2}{\alpha}}-\rho_{\!D}^{\frac{2}{\alpha}}}\right)^2.
\end{align}
\else
\begin{small}
\begin{align}
\left(x-\frac{\xi^{\frac{2}{\alpha}}\dis}{\xi^{\frac{2}{\alpha}}-\rho_{\!D}^{\frac{2}{\alpha}}}\right)^2+y^2<\left(\frac{\xi^{\frac{1}{\alpha}}\rho_{\!D}^{\frac{1}{\alpha}}\dis}{\xi^{\frac{2}{\alpha}}-\rho_{\!D}^{\frac{2}{\alpha}}}\right)^2.
\end{align}
\end{small}
\fi
The above expression can be interpreted as follows: if p-DUE is inside a disk region centered at $\left(\frac{\xi^{\frac{2}{\alpha}}\dis}{\xi^{\frac{2}{\alpha}}-\rho_{\!D}^{\frac{2}{\alpha}}},0\right)$ with radius $\frac{\xi^{\frac{1}{\alpha}}\rho_{\!D}^{\frac{1}{\alpha}}\dis}{\xi^{\frac{2}{\alpha}}-\rho_{\!D}^{\frac{2}{\alpha}}}$, this P-DUE is not in D2D mode. Moreover, since the p-DUE is always surrounding around its DRx, the p-DUE is confined within the disk region centered at origin with radius $\rd$. Combining these two requirements, we obtain that when the p-DUE is inside the overlap region of these two disk regions (i.e., the shaded area in Fig.~\ref{fig:active2}, denoted as $\mathcal{B}_2$), this p-DUE is not in D2D mode. Hence, we have the probability that a p-DUE is in D2D mode is $1-\frac{\mathcal{B}_2}{\pi\rd^2}$. Note that $\mathcal{B}_2=
\psi\!\left(\!\frac{\xi^{\frac{2}{\alpha}}\dis}{\xi^{\frac{2}{\alpha}}-\rho_{\!D}^{\frac{2}{\alpha}}},
             \rd,\frac{\xi^{\frac{1}{\alpha}}\rho_{\!D}^{\frac{1}{\alpha}}\dis}{\xi^{\frac{2}{\alpha}}-\rho_{\!D}^{\frac{2}{\alpha}}}\!\right)$ if $\dis\!\geq\!\left(\!1\!-\!\left(\!\frac{\rho_{\!D}}{\xi}\right)^{\!\frac{1}{\alpha}}\!\right)\!\rd$, while $\mathcal{B}_2=
\pi\left(\frac{\xi^{\frac{1}{\alpha}}\dis}{\xi^{\frac{1}{\alpha}}-\rho_{\!D}^{\frac{1}{\alpha}}}\right)^2$ for $\dis\!<\!\left(\!1\!-\!\left(\!\frac{\rho_{\!D}}{\xi}\right)^{\!\frac{1}{\alpha}}\!\right)\!\rd$, where $\psi(\cdot,\cdot,\cdot)$ is defined in~\eqref{eq:psi} in Lemma 1.

Likewise, we can derive $p_{\DD}(\dis)$ for $\xi\leq\rho_{\!D}$ using the same approach. Due to the space limitation, we do not present the derivation here.
\subsection{Different path-loss exponent}
 We can directly write the probability of being in D2D mode as
 \ifCLASSOPTIONonecolumn
\begin{align}
p_{\DD}(\dis)=&\Pr\left(\rho_{\!D} r_{d}^{\alpha_{\!D}}r_{c}^{-\alpha_{\!C}}<\xi \right)\overset{(a)}{\approx}\Pr\left(h<\frac{\xi r_c^{\alpha_{\!C}}}{\rho_{\!D}r_{d}^{\alpha_{\!D}}}\right)\nonumber\\
=&\Pr\left(h<\frac{\xi \left(r_d^2+\dis^2-2r_d\dis\cos(\theta)\right)^{\alpha_{\!C}/2}}{\rho_{\!D}r_{d}^{\alpha_{\!D}}}\right)\nonumber\\
\overset{(b)}{\approx}&\mathbb{E}_{r_d,\theta}\left\{\left(1-\exp\left(-\frac{N\xi \left(r_d^2+\dis^2-2r_d\dis\cos(\theta)\right)^{\alpha_{\!C}/2}}{(N!)^{1/N}\rho_{\!D}r_{d}^{\alpha_{\!D}}}\right)\right)^N\right\},
\end{align}
\else
\begin{align}
&p_{\DD}(\dis)=\Pr\left(\rho_{\!D} r_{d}^{\alpha_{\!D}}r_{c}^{-\alpha_{\!C}}<\xi \right)\overset{(a)}{\approx}\Pr\left(h<\frac{\xi r_c^{\alpha_{\!C}}}{\rho_{\!D}r_{d}^{\alpha_{\!D}}}\right)\nonumber\\
&=\!\Pr\!\left(h<\frac{\xi \left(r_d^2+\dis^2-2r_d\dis\cos(\theta)\right)^{\alpha_{\!C}/2}}{\rho_{\!D}r_{d}^{\alpha_{\!D}}}\right)\nonumber\\
&\overset{(b)}{\approx}\!\mathbb{E}_{r_d,\theta}\!\left\{\!\!\left(\!1-\exp\!\left(\!-\frac{N\xi \left(r_d^2\!+\!\dis^2\!-\!2r_d\dis\cos(\theta)\right)^{\alpha_{\!C}/2}}{(N!)^{1/N}\rho_{\!D}r_{d}^{\alpha_{\!D}}}\!\right)\!\right)^{\!\!N}\!\right\},
\end{align}
\fi
\noindent where $(a)$ comes from the introduction of a dummy random variable $h$, which follows the Gamma distribution with parameter $N$, and the fact the normalized Gamma distribution
converges to identity when its parameter goes to infinity~\cite{Bai-2015}, and $(b)$ comes from the approximation of a Gamma distribution~\cite{Alzer-1997}.

It is not easy to find the closed-form result for this probability. Instead, we consider an approximation, in which the distance between BS and p-DUE $r_c$ is approximated by the distance between BS and DRx $\dis$. Hence, by substituting $\sqrt{r_d^2+\dis^2-2r_d\dis\cos(\theta)}$ by $\dis$ in the above expression and using the Binomial theorem, we get
\ifCLASSOPTIONonecolumn
\begin{align}
p_{\DD}(\dis)&\approx1+\sum_{n=1}^{N}\left(-1\right)^{n}\binom {N}{n}\int_0^{\rd}\left(1-\exp\left(-\frac{nN\xi \dis^{\alpha_{\!C}}}{(N!)^{1/N}\rho_{\!D}r_{d}^{\alpha_{\!D}}}\right)\right)\frac{2r_d}{\rd^2}\,\textup{d}r_d\nonumber\\
&=1+\sum_{n=1}^{N}\left(-1\right)^{n}\binom {N}{n}\frac{2\dis^{2\frac{\alpha_{\!C}}{\alpha_{\!D}}}\left(nN\xi\right)^{\frac{2}{\alpha_{\!D}}}\Gamma\!\left[-\frac{2}{\alpha_{\!D}},\frac{\dis^{\alpha_{\!C}}nN\xi}{(N!)^{1/N}\rho_{\!D}\rd^{\alpha_{\!D}}}\right]}{\rd^2\alpha_{\!D}\left((N!)^{1/N}\rho_{\!D}\right)^{\frac{2}{\alpha_{\!D}}}}.
\end{align}
\else
\begin{align}
p_{\DD}(\dis)&\approx1+\sum_{n=1}^{N}\left(-1\right)^{n}\binom {N}{n}\nonumber\\
&\quad\times\int_0^{\rd}\left(1-\exp\left(-\frac{nN\xi \dis^{\alpha_{\!C}}}{(N!)^{1/N}\rho_{\!D}r_{d}^{\alpha_{\!D}}}\right)\right)\frac{2r_d}{\rd^2}\,\textup{d}r_d\nonumber\\
&=1+\sum_{n=1}^{N}\left(-1\right)^{n}\binom {N}{n}\frac{2\dis^{2\frac{\alpha_{\!C}}{\alpha_{\!D}}}\left(nN\xi\right)^{\frac{2}{\alpha_{\!D}}}\!}{\rd^2\alpha_{\!D}\left((N!)^{1/N}\rho_{\!D}\right)^{\frac{2}{\alpha_{\!D}}}}\nonumber\\
&\quad\quad\quad\quad\quad\times \Gamma\left[-\frac{2}{\alpha_{\!D}},\frac{\dis^{\alpha_{\!C}}nN\xi}{(N!)^{1/N}\rho_{\!D}\rd^{\alpha_{\!D}}}\right].
\end{align}
\fi

Thus, we obtain the probability of being in D2D mode in Proposition~\ref{prop_active}.
\end{proof}

\end{document}